\documentclass[12pt,preprint]{aastex}
\usepackage{epsf}

\def\linebreak{\hfil\break}

\def\
{\hfil\linebreak}

\def\deg{\ifmmode {^\circ}\else {$^\circ$}\fi}
\def\degree{\ifmmode {^\circ}\else {$^\circ$}\fi}
\def\mum{\ifmmode {\rm \,\mu {\rm m}}\else $\rm \,\mu {\rm m}$\fi}
\def\arcsec{\ifmmode ^{\prime \prime}\else $^{\prime \prime}$\fi}

\def\inch{\ifmmode ^{\prime \prime}\else $^{\prime \prime}$\fi}
\def\mjup{\ifmmode { M_J}\else $ M_J$\fi}
\def\mjupyr{\ifmmode {M_J~{\rm yr^{-1}}}\else $M_J~{\rm yr^{-1}}$\fi}
\def\rjup{\ifmmode { R_J}\else $ R_J$\fi}
\def\mearth{\ifmmode { M_{\oplus}}\else $ M_{\oplus}$\fi}
\def\rearth{\ifmmode { R_{\oplus}}\else $ R_{\oplus}$\fi}
\def\msunyr{\ifmmode {M_{\odot}~{\rm yr^{-1}}}\else $M_{\odot}~{\rm yr^{-1}}$\fi}
\def\msun{\ifmmode {M_{\odot}}\else $M_{\odot}$\fi}
\def\rsun{\ifmmode {R_{\odot}}\else $R_{\odot}$\fi}
\def\lsun{\ifmmode {L_{\odot}}\else $L_{\odot}$\fi}
\def\mstar{\ifmmode {M_{\star}}\else $M_{\star}$\fi}
\def\rstar{\ifmmode {R_{\star}}\else $R_{\star}$\fi}
\def\tstar{\ifmmode {T_{\star}}\else $T_{\star}$\fi}
\def\lstar{\ifmmode {L_{\star}}\else $L_{\star}$\fi}
\def\ldlstar{\ifmmode { L_d / L_{\star}}\else $ L_d / L_{\star}$\fi}
\def\ad{\ifmmode {A_d}\else $A_d$\fi}
\def\md{\ifmmode {M_d}\else $M_d$\fi}
\def\ld{\ifmmode {L_d}\else $L_d$\fi}
\def\qdstar{\ifmmode Q_D^\star\else $Q_D^\star$\fi}
\def\resc{\ifmmode r_{esc}\else $r_{esc}$\fi}
\def\mesc{\ifmmode m_{esc}\else $m_{esc}$\fi}
\def\rmin{\ifmmode r_{min}\else $r_{min}$\fi}
\def\rmax{\ifmmode r_{max}\else $r_{max}$\fi}
\def\mmin{\ifmmode m_{min}\else $m_{min}$\fi}
\def\mmax{\ifmmode m_{max}\else $m_{max}$\fi}
\def\rmind{\ifmmode r_{min,d}\else $r_{min,d}$\fi}
\def\rmaxd{\ifmmode r_{max,d}\else $r_{max,d}$\fi}
\def\mmaxd{\ifmmode m_{max,d}\else $m_{max,d}$\fi}
\def\vrad{\ifmmode v_{rad}\else $v_{rad}$\fi}
\def\qz{\ifmmode q_{0}\else $q_{0}$\fi}
\def\qi{\ifmmode q_{i}\else $q_{i}$\fi}
\def\ql{\ifmmode q_{l}\else $q_{l}$\fi}
\def\qs{\ifmmode q_{s}\else $q_{s}$\fi}
\def\bd{\ifmmode b_{d}\else $b_{d}$\fi}
\def\bl{\ifmmode b_{L}\else $b_{L}$\fi}
\def\2470{[24]--[70]}

\def\rhill{\ifmmode R_H\else $R_H$\fi}
\def\rfast{\ifmmode R_{fast}\else $R_{fast}$\fi}
\def\rgap{\ifmmode R_{gap}\else $R_{gap}$\fi}
\def\vhill{\ifmmode v_H\else $v_H$\fi}
\def\rbrk{\ifmmode r_{brk}\else $r_{brk}$\fi}
\def\rdamp{\ifmmode r_{damp}\else $r_{damp}$\fi}
\def\rin{\ifmmode r_{in}\else $r_{in}$\fi}
\def\rout{\ifmmode r_{out}\else $r_{out}$\fi}
\def\tin{\ifmmode t_{in}\else $t_{in}$\fi}
\def\tout{\ifmmode t_{out}\else $t_{out}$\fi}
\def\ain{\ifmmode {a_{in}}\else $a_{in}$\fi}
\def\amax{\ifmmode {a_{max}}\else $a_{max}$\fi}
\def\aout{\ifmmode {a_{out}}\else $a_{out}$\fi}
\def\ml0{\ifmmode m_{L,0}\else $m_{L,0}$\fi}
\def\f0{\ifmmode f_{0}\else $f_{0}$\fi}
\def\r0{\ifmmode r_{0}\else $r_{0}$\fi}
\def\R0{\ifmmode R_{0}\else $R_{0}$\fi}
\def\m0{\ifmmode m_{0}\else $m_{0}$\fi}
\def\M0{\ifmmode M_{0}\else $M_{0}$\fi}
\def\xm{\ifmmode x_{m}\else $x_{m}$\fi}
\def\sigz{\ifmmode \Sigma_0\else $\Sigma_0$\fi}
\def\ergg{\ifmmode {\rm erg~g^{-1}}\else $\rm erg~g^{-1}$\fi}
\def\gyr{\ifmmode {\rm g~yr^{-1}}\else ${\rm g~yr^{-1}}$\fi}
\def\cms{\ifmmode {\rm cm~s^{-1}}\else ${\rm cm~s^{-1}}$\fi}
\def\kms{\ifmmode {\rm km~s^{-1}}\else $\rm km~s^{-1}$\fi}
\def\ms{\ifmmode {\rm m~s^{-1}}\else $\rm m~s^{-1}$\fi}
\def\gcms{\ifmmode {\rm g~cm^{-2}}\else $\rm g~cm^{-2}$\fi}
\def\gcmc{\ifmmode {\rm g~cm^{-3}}\else $\rm g~cm^{-3}$\fi}
\def\atil{\ifmmode {\tilde{a}}\else $\tilde{a}$\fi}
\def\ttil{\ifmmode {\tilde{t}}\else $\tilde{t}$\fi}
\def\sqrttt{\ifmmode {\tilde{t}^{1/2}}\else $\tilde{t}^{1/2}$\fi}

\def\VP{2012 VP$_{113}$}

\begin{document}

\title{Making Planet Nine: Pebble Accretion at 250--750 AU in a Gravitationally Unstable Ring}
\vskip 7ex
\author{Scott J. Kenyon}
\affil{Smithsonian Astrophysical Observatory,
60 Garden Street, Cambridge, MA 02138} 
\email{e-mail: skenyon@cfa.harvard.edu}

\author{Benjamin C. Bromley}
\affil{Department of Physics \& Astronomy, University of Utah, 
201 JFB, Salt Lake City, UT 84112} 
\email{e-mail: bromley@physics.utah.edu}
%
%

\begin{abstract}

We investigate the formation of icy super-Earth mass planets within a 
gravitationally unstable ring of solids orbiting at 250--750~AU around 
a 1~\msun\ star. Coagulation calculations demonstrate that a system of 
a few large oligarchs and a swarm of pebbles generates a super-Earth
within 100--200~Myr at 250~AU and within 1--2~Gyr at 750~AU. Systems 
with more than ten oligarchs fail to yield super-Earths over the age 
of the solar system. As these systems evolve, destructive collisions
produce detectable debris disks with luminosities of 
$10^{-5} - 10^{-3}$ relative to the central star.
\end{abstract}

\keywords{planetary systems -- planets and satellites: formation -- 
solar system: formation }

\section{INTRODUCTION}
\label{sec: intro}

Over the past 10--15 yr, the discoveries of Sedna, \VP, and other
dwarf planets have renewed interest in the architecture of the 
outer solar system \citep[e.g.,][and references therein]{brown2004,
sheppard2010a,chen2013,trujillo2014}. Today, several fairly large 
($R \approx$ 200--1000~km) dwarf planets are known to have orbits 
with semimajor axis $a \gtrsim$ 150~AU, eccentricity $e \gtrsim$ 0.7, 
perihelion distance $q_p \ge$ 30~AU, argument of perihelion 
$\omega \approx$ 310\deg, and longitude of perihelion
$\varpi \approx$ 71\deg.  The observed distributions of $\omega$ 
and $\varpi$ for these objects are statistically unlikely 
\citep[e.g.,][]{trujillo2014,delafm2014,batygin2016}. 

Although dynamical interactions between the Sun and a passing star 
can produce objects on highly eccentric orbits like Sedna and 
\VP\ \citep[e.g.,][]{morby2004a,kb2004d,jilkova2015,li2016}, 
torques from Jupiter and the other gas giants randomize $\omega$ 
and $\varpi$ on 0.1--1~Gyr time scales 
\citep{gomes2006,lykaw2008,trujillo2014,batygin2016}.
A super-Earth mass planet at $a \approx$ 200--1000~AU can maintain
the observed distributions of $\omega$ and $\varpi$ for the age of 
the solar system \citep{trujillo2014,delafm2014,iorio2014,batygin2016,
brown2016}. 
This planet might also generate a set of trans-Neptunian objects with
high orbital inclination \citep[e.g.,][]{gladman2009,rabin2013},
account for the properties of some comets \citep[e.g.,][]{matese1999,
gomes2006,lykaw2008}, and perhaps improve the residuals in model fits 
to the orbit of Saturn \citep[e.g.,][and references therein]{fienga2016}.

To explore the origin of a possible `planet nine', we have previously 
considered two broad options \citep{bk2014,kb2015a}. In one mechanism, 
the protoplanets destined to become Jupiter, Saturn, Uranus, and Neptune 
scatter a lower mass, icy protoplanet from 5--15~AU to several hundred 
AU. Interactions with the gaseous disk then circularize the orbit. For 
plausible initial conditions, super-Earth mass planets achieve orbits 
similar to that proposed for planet nine. In another picture, the gaseous 
disk leaves behind a ring of solid material at 100--250~AU. Collisional 
evolution within the ring produces super-Earth mass planets on 1--2~Gyr 
time scales.  In principle, the orbital parameters of planet nine can 
distinguish between these two options.

Aside from applications to the outer solar system, these results have
broad implications for the architectures of exoplanetary systems.  The
process of scattering, orbit circularization, and subsequent growth of 
icy protoplanets at $a \gtrsim$ 20--30~AU plausibly explains the orbits 
of the gas giants in HR~8799 \citep{marois2008}, 1RXS J160929.1$-$210524 
\citep{lafren2010}, and HD~106906~b \citep{bailey2014}.  Generation of 
dust grains during {\it in situ} formation of a super-Earth at 
100--250~AU can account for the debris disks in HD~107146 
\citep{corder2009,ricci2015}, HD~202628 \citep{krist2012}, and 
HD~207129 \citep{krist2010,marshall2011}. 

To improve predictions for super-Earth formation at 250--750~AU, we 
consider an expanded set of calculations for {\it in situ} formation 
of planet nine. After developing the general picture for the origin 
of a ring of solids at $a \gtrsim$ 100~AU in \S\ref{sec: back}, we 
outline the numerical method (\S\ref{sec: calcs}) and summarize the 
major results (\S\ref{sec: evol}). We conclude with a discussion 
(\S\ref{sec: disc}) and brief summary (\S\ref{sec: summ}).

\section{BACKGROUND}
\label{sec: back}

Opaque protoplanetary disks surround all newly-formed stars \citep[][and 
references therein]{kh1995,kgw2008,will2011,andrews2015,tobin2015}.  Most 
disks have radii $R_d \approx$ 10--1000~AU and enough solid material to 
build a typical planetary system \citep[see also][]{najita2014}.  On time 
scales of 1--10~Myr, the optically thick disk disappears \citep{haisch2001,
mama2009,will2011,alex2014}; accretion from the disk onto the central 
star also ceases \citep{hart1998}.

Current observations are consistent with several plausible paths for disk 
disappearance \citep[e.g.,][and references therein]{najita2007b,currie2009,
cieza2010,espaillat2010,andrews2011,kim2013,najita2015}. In some systems, 
large inner holes suggest dissipation from the inside out, as predicted 
by theoretical models of photoevaporating disks \citep[e.g.,][]{owen2012,
gorti2015}. In others, material may vanish roughly simultaneously throughout 
the disk.

Forming super-Earth mass planets {\it in situ} at 100--1000~AU requires 
a large amount of solid material. Although gaseous disks often extend to 
500--1000~AU, the solids in the youngest disks rarely extend beyond 
50--100~AU \citep[][and references therein]{birnstiel2014,tobin2015,
andrews2015,canovas2016}. Thus, we seek a mechanism to transport small 
particles radially outward. Although developing a robust calculation of
radial transport is beyond the scope of this paper, we show that gas
drag in a photoevaporating protoplanetary disk is a plausible mechanism 
for generating a ring of pebbles beyond 100~AU.

\subsection{Radial transport of small particles in protoplanetary disks}

In all protoplanetary disks, pressure causes the gas to orbit the central 
star somewhat more slowly than the local circular velocity \citep{ada76,
weiden1977a}.  Solid particles then feel a headwind which drags them 
towards local pressure maxima \citep[see also][]{youdin2002,youdin2004a,
brauer2008,birn2010,chiang2010,windmark2012,garaud2013}.  
For 1--100~cm particles, the drift velocity is roughly 0.1\% of the 
orbital velocity.  Thus, drift times are short.

When the disk has a smooth radial surface density gradient, 
$\Sigma \propto a^{-p}$ with $p > 0$, small solids typically drift 
radially inward \citep[e.g.,][and references therein]{raf2004,
youdin2004a,chiang2010}.  As small particles spiral in toward the 
central star, they may concentrate within local pressure maxima 
produced by vortices or dust traps \citep[e.g.,][]{klahr1997,
chavanis2000,haghi2003,heng2010b,raettig2015,gibbons2015} or 
at the edges of gaps produced by massive planets 
\citep{ayliffe2012,zhu2014,gonzalez2015}.

This behavior suggests a mechanism for outward drift of small particles.
Consider a disk with inner edge \ain\ and outer edge \aout\ orbiting a
central star with radius \rstar. If $\ain \gg \rstar$ and $\aout \gg \ain$,
the disk surface density $\Sigma$ rises from zero at \ain\ to some maximum 
$\Sigma_{max}$ and then declines monotonically outward.  If the disk 
temperature also follows a power-law, $T \propto r^{-q}$, the disk 
has a maximum pressure $P_{max}$ at some $a$ = \amax\ close to \ain.  
Small particles at $\amax\ \le a \le \ain$ ($a \ge \amax$) then drift 
radially outward (inward) until they reach \amax. Now suppose \ain\ and 
\amax\ expand on a time scale longer than the radial drift time. If the 
disk maintains a pressure maximum at \amax, particles are continuously 
swept from \ain\ to \amax. Once the disk dissipates, it leaves behind 
a ring of small particles at the last \amax. A sufficiently large final 
\amax\ enables super-Earth formation at 250--750~AU.

Photoevaporating disks offer one way to achieve this evolution. In 
theoretical models \citep[e.g.,][and references therein]{owen2012,
gorti2015}, high energy photons from the central star and the inner disk
ionize material above the disk photosphere. The high temperature of this 
material generates a wind which removes gas from the disk. As the system 
evolves, the surface density of the disk declines. Eventually, mass loss
creates an inner hole devoid of gas. Expansion of the inner hole gradually
removes more and more material from the disk until the disk vanishes.

For plausible protostellar disks, the expansion rate of the inner edge of
the disk is much smaller than the radial drift rate of small particles.  
Numerical calculations \citep{clarke2001,gorti2009,owen2012,owen2013,
gorti2015,haworth2016} and observations \citep{calvet2005,currie2009,
cieza2010,andrews2011,najita2015} indicate an expansion rate 
$\dot{a}_{in} \gtrsim$ 10--20~AU~Myr$^{-1}$ $\approx$ 5--10~\cms.  
Typical radial drift rates for 1--10~cm particles are 3--30~\ms.  
For photoevaporation models with $\dot{a}_{in} \lesssim$ 
1000--2000~AU~Myr$^{-1}$, it seems plausible that the expanding inner 
edge of the disk can transport small particles to large radial distances.

\subsection{Dynamical Cooling and Gravitational Instability}

As the inner edge of the disk expands, we assume that turbulence within the
gas prevents swept up small particles from colliding and merging into larger
objects. Once the gas has dissipated, turbulence is minimal. In \citet{kb2015a}, 
we considered the collisional evolution of sets of mono-disperse distributions 
of particles covering a range of sizes, 1~cm to $10^3$~km. Here, we explore 
outcomes when a collection of cm-sized `pebbles' becomes gravitationally 
unstable and produces much larger oligarchs.

In a swarm of pebbles with velocity dispersion $c_p$, surface density $\Sigma$, 
and angular velocity $\Omega$, the system is gravitationally unstable when 
$c_p \Omega < \pi G \Sigma$ \citep[e.g.,][]{chiang2010}.  Setting 
$\Sigma = \Sigma_0 (a / {\rm 1~AU})^{-3/2}$, gravitationally stability 
requires a particle eccentricity $e \gtrsim e_s$,
where
\begin{equation}
e_s \approx 10^{-4} \left ( { \Sigma_0 \over {\rm 30~g~cm^{-2}} } \right ) \left ( {a \over {\rm 100~AU} } \right )^{1/2} ~ .
\label{eq: e-grav}
\end{equation}
For material at 250--750~AU, the minimum $e$ for gravitational stability 
is $e_s \approx 1-3 \times 10^{-4}$.

Once the disk dissipates, pebbles with an initial $e \approx 10^{-3}$ set 
by disk turbulence damp very quickly. In \citet{kb2015a}, the e-folding time 
for collisional damping to reduce $e$ is roughly $10^4$~yr at 125~AU. The
damping time, $t_d \propto P / \Sigma$, scales with $\Sigma$ and the orbital 
period $P$. 
For annuli with identical total masses in pebbles and width $\delta a = 0.2 a$,
the damping time is roughly $10^5$ yr at 250~AU and $5 \times 10^6$ yr at
750~AU. These time scales are reasonably short, so we assume the pebbles 
damp into a gravitationally unstable ring.

Outcomes of gravitational instability remain uncertain \citep[e.g.,][and 
references therein]{michi2007,chiang2010,johan2015,simon2015}. Here, we 
follow \citet{michi2009,michi2010} and assume that the instability produces 
one or more oligarchs with a size \rmax\ set by the wavelength of the shortest 
stable mode in the swarm of pebbles. For the conditions we consider below, 
\rmax\ $\approx$ 100~km. 
Once $N$ oligarchs collapse out of the swarm, they begin to accrete leftover 
pebbles. At the same time, pebbles begin to collide and merge into larger 
objects. Although this set of starting conditions is somewhat artificial, it 
captures the spirit of the likely evolution following gravitational instability
in a ring of pebbles.

\section{PLANET FORMATION CALCULATIONS}
\label{sec: calcs}

To investigate planet growth at 250--750~AU, we consider coagulation
calculations for a single annulus with width $\delta a$ at semimajor 
axis $a$ \citep[][see Table 1 for a list of parameters]{kb2015a}. 
Solid material orbits a central star with mass \mstar\ = 1~\msun. 
The annulus contains $M$ mass batches (labeled from $k$ = 1 to $k$ = $M$)
with characteristic mass $m_k$ and radius $r_k$ \citep{weth1993,kl1998}. 
Batches are logarithmically spaced in mass, with mass ratio 
$\delta \equiv m_{k+1} / m_{k}$.  Each mass batch contains $N_k$ 
particles with total mass $M_k$ and average mass $\bar{m}_k = M_k / N_k$. 
Particle numbers $N_k < 10^{15}$ are always integers.  Throughout the 
calculation, various algorithms use the average mass to calculate the 
average physical radius $\bar{r}_k$, collision cross-section, collision 
energy, and other necessary physical variables.  As mass is added and 
removed from each batch, the number of particles, the total mass, and
the average mass change \citep{weth1993,kl1998,kb2015a,kb2016}.

In these calculations, we follow the evolution of particles with mass 
density $\rho_p$ = 1.5~g~cm$^{-3}$ and sizes ranging from a minimum 
\rmin\ = 1~\mum\ to a maximum \rmax. The mass spacing factor 
$\delta = 2^{1/4}$.
Initially, the annulus contains two mono-disperse swarms of planetesimals 
with initial sizes \r0\ and \rmax, total surface density $\Sigma_0$, total 
mass $M_0$, and horizontal and vertical velocities $h_0$ and $v_0$ relative 
to a circular orbit.  Within the swarm, the large particles contain a
fraction $f_0$ of the initial mass $M_0$.  The horizontal velocity depends 
on the orbital eccentricity, $e$ = 1.6 $(h/v_K)^2$, where $v_K$ is the 
circular orbital velocity.  The orbital inclination is ${\rm sin}~i$ = 
$\sqrt{2} v/v_K$.  

The mass and velocity distributions of the planetesimals evolve in time 
due to inelastic collisions, drag forces, and gravitational encounters.  
This evolution depends on the solution to a coupled set of coagulation 
and Fokker-Planck
equations which treats the outcomes of mutual collisions between all 
particles in all mass bins \citep{kl1998,kl1999a,kb2004a,kb2008,kb2015a,
kb2016}.  For physical collisions, we adopt the particle-in-a-box algorithm; 
the collision rate is then $n \sigma v f_g$, where $n$ is the number density 
of objects, $\sigma$ is the geometric cross-section, $v$ is the relative 
velocity, and $f_g$ is the gravitational focusing factor \citep{weth1993,kl1998}.  
For a specific mass bin, the solutions include terms for (i) loss of mass from 
mergers with other objects and (ii) gain of mass from collisional debris and 
mergers of smaller objects.

Depending on physical conditions in the disk, we derive $f_g$ in the 
dispersion or the shear regime \citep{kl1998,gold2004,kb2012,youdin2013}.  
To set conditions in the shear regime, we define the Hill radius
\begin{equation}
\rhill = a (m / 3 ~ \mstar)^{1/3} ~ .
\end{equation}
When interpreting results of the calculations, it is useful to track 
the Hill radius of the largest object with $m = \mmax$ and the relative 
eccentricity, $e_{rel} = e a / \rhill$, of small particles. When
$e_{rel} \lesssim$ 1 ($\gtrsim$ 1), collisions with the largest object
are in the shear (dispersion) regime.

For the most massive super-Earths produced in these simulations, the Hill
radius is $R_H \approx 0.02 a$. With the annulus width $\delta a$ = $0.2 a$,
the width of the annulus is much larger than the Hill radius of the largest
objects in the simulation.

Within the single annulus, the most massive protoplanets on roughly circular 
orbits are `isolated' from one another \citep{weth1993}. Isolated protoplanets 
can accrete smaller objects but cannot collide with other isolated protoplanets. 
Our algorithm identifies the $n$ ($n+1$) largest objects whose combined 
gravitational range is smaller (larger) than $\delta a$ and establishes 
these $n$ objects as isolated protoplanets \citep{kl1998,kb2015a}. 

Collision outcomes depend on the ratio $Q_c/\qdstar$, where $\qdstar$ is 
the collision energy needed to eject half the mass of a pair of colliding 
planetesimals to infinity and $Q_c$ is the center of mass collision energy 
\citep[see also][]{weth1993,will1994,tanaka1996b,stcol1997a,kl1999a,obrien2003,
koba2010a}.  A colliding pair of planetesimals with horizontal velocity 
$h_1$, $h_2$ and vertical velocity $v_1$, $v_2$ have relative horizontal and 
vertical velocities $h_c = (h_1^2 + h_2^2)^{1/2}$ and $v_c = (v_1^2 + v_2^2)^{1/2}$ 
\citep[see also][]{weth1993,kl1998,kb2004a,kb2015a}. The escape velocity of 
the colliding pair is $v_{esc} = (2 G m_c / r_c)^{1/2}$, where $m_c = m_1 + m_2$ 
is the combined mass and $r_c = r_1 + r_2$ is the combined radius. The center of 
mass collision energy is then
\begin{equation}
Q_c = 0.5 \mu (h_c^2 + v_c^2 + v_{esc}^2) / m_c
\label{eq: Qc}
\end{equation}
where $\mu = m_1 m_2 / m_c $ is the reduced mass.

Consistent with N-body simulations of collision outcomes 
\citep[e.g.,][]{benz1999,lein2008,lein2009}, we set
\begin{equation}
\qdstar = Q_b r^{\beta_b} + Q_g \rho_p r^{\beta_g}
\label{eq: Qd}
\end{equation}
where $Q_b r^{\beta_b}$ is the bulk component of the binding energy and
$Q_g \rho_p r^{\beta_g}$ is the gravity component of the binding energy.
For `strong' planetesimals, 
we adopt $Q_b$ = $2 \times 10^5$ erg g$^{-1}$ cm$^{0.4}$, 
$\beta_b = -0.40$, $Q_g$ = 0.22 erg g$^{-2}$ cm$^{1.7}$, 
and $\beta_g$ = 1.30.  These parameters are broadly consistent with 
published analytic and numerical simulations \citep[e.g.,][]{davis1985,
hols1994,love1996,housen1999}.  At small sizes, they agree with results 
from laboratory \citep[e.g.,][]{ryan1999,arakawa2002,giblin2004,burchell2005}
and numerical \citep[e.g.,][]{lein2009} experiments of impacts between icy 
objects.  For $r \approx$ 10--100~m particles with the smallest \qdstar, 
\qdstar\ is a factor of 3--10 smaller than in other studies 
\citep[e.g.,][]{bottke2010}.  Thus, these small planetesimals are 
relatively easy to break.

In some calculations, we consider ensembles of `weak' planetesimals with 
$Q_b = 10^3$~erg~g$^{-1}$ and $\beta_b$ = 0. Among these objects, small 
objects have negligible material strength; the binding energies of large 
objects are set by gravity as for strong planetesimals.

For two colliding planetesimals with masses $m_1$ and $m_2$, the mass of the 
merged planetesimal is
\begin{equation}
m = m_1 + m_2 - \mesc ~ ,
\label{eq: msum}
\end{equation}
where the mass of debris ejected in a collision is
\begin{equation}
\mesc = 0.5 ~ (m_1 + m_2) \left ( \frac{Q_c}{Q_D^*} \right)^{b_d} ~ .
\label{eq: mej}
\end{equation}
The exponent $b_d$ is a constant of order unity 
\citep[e.g.,][]{davis1985,weth1993,kl1999a,benz1999,obrien2003,lein2012}. 
We adopt $b_d$ = 1 \citep[see also][]{koba2010a,koba2010b,kb2015a}. 

To place the debris in the grid of mass bins, we set the mass of the largest 
collision fragment as 
\begin{equation}
\mmaxd = {\rm min} (\ml0, \ml0 ~ (Q_c / \qdstar)^{-\bl})~\mesc ~
\label{eq: mmaxd}
\end{equation}
and adopt a differential size distribution $N(r) \propto r^{-3.5}$. 
After placing a single object with mass \mmaxd\ in the grid, we place 
material in successively smaller mass bins until (i) the mass is 
exhausted or (ii) mass is placed in the smallest mass bin. Any 
material left over is removed from the grid. For these calculations,
we adopt \ml0\ = 0.2 and \bl\ = 0 or 1. 

As we place the debris in specific mass bins, we also redistribute the
kinetic energy per unit mass of each colliding pair of planetesimals. 
As in \citet{kl1998}, we assume all collisions between mass batches 
conserve the horizontal and vertical components of kinetic energy. For
an initial kinetic energy, $m_1 (h_1^2 + v_1^2) + m_2 (h_2^2 + v_2^2)$, 
any merged planetesimal with mass $m$ receives a fraction $m / (m_1 + m_2)$
of this kinetic energy; any fragment with mass $m_f$ receives a fraction
$m_f / (m_1 + m_2)$.  Recalling the center of mass collision energy from 
eq. (\ref{eq: Qc}), this approach assumes that the escape velocity
component of the collision energy is equal to the energy required to 
disperse the fragments to infinity. 

To compute the evolution of the velocity distribution, we also include 
collisional damping from inelastic collisions and gravitational 
interactions.  For inelastic and elastic collisions, we follow the 
statistical, Fokker-Planck approaches of \citet{oht1999} and 
\citet{oht2002}, which treat pairwise interactions (e.g., dynamical 
friction and viscous stirring) between all objects. For evaluating 
these interactions within a single annulus, we eliminate terms to
calculate the probability that objects in one annulus interact with 
objects in other annuli \citep{kb2001,kb2004b,kb2008}. We also compute 
long-range stirring from distant oligarchs \citep{weiden1989}. 
At 250--750~AU, particles collide and grow on time scales much longer 
than the 1--5~Myr lifetime of the gaseous disk \citep{haisch2001,will2011,
clout2014}.  Thus, we set the initial surface density of the gas to zero 
and ignore gas drag on small solids \citep{ada76,weiden1977a,raf2004}. 

In previous studies, we have compared results from the coagulation code 
with analytical models and other numerical simulations of collision 
rates/outcomes, collisional damping, and gravitational stirring
\citep{kl1998,kb2001,kb2004c,kb2006,bk2006,kb2008,kb2015a,kb2016}.  Our 
calculations yield excellent matches to these results.  Typical solutions 
conserve mass and energy to machine accuracy.  Over the $10^6$ timesteps 
in a typical 10~Gyr run, mass and energy are conserved to better than a 
part in $10^{10}$.

\section{EVOLUTION OF THE LARGEST OBJECTS}
\label{sec: evol}

To evolve a sea of planetesimals in a single annulus, we set the sizes 
\r0\ and \rmax, the surface density $\Sigma_0$, the fraction \f0\ of 
the initial mass in the largest objects, and the orbital elements $e_0$ 
and $i_0$.  For these simulations, \r0\ = 1~cm, \rmax\ = 100~km, and
\f0\ = 0.5 to $10^{-7}$. The total mass in the annulus is 15.8~\mearth.    
The surface density of an annulus with $\delta a = 0.2 a$ is then 
$\Sigma_0$ = $5.4 \times 10^{-3}$~\gcms\ at 250~AU 
($6 \times 10^{-4}$~\gcms\ at 750~AU). For comparison, the minimum mass 
solar nebula has $\Sigma \approx$ 30~g~cm$^{-2}~(a/{\rm 1~AU})^{-3/2}$ 
\citep{kb2008}, which implies $\Sigma$ = $7.6 \times 10^{-3}$~g~cm$^{-2}$ 
at $a$ = 250~AU ($\Sigma$ = $1.5 \times 10^{-3}$~g~cm$^{-2}$ at $a$ 
= 750~AU). Evolution times depend inversely on the mass in solid 
objects \citep{kl1999a,kb2002b,kb2008,kb2010}; thus, we consider only 
one surface density at each $a$.

In most published simulations, the initial orbital elements of 
planetesimals are set to match the escape velocity of the largest 
objects \citep[e.g.,][]{koba2010b,weid2010}.  For particles with 
\r0\ $\gtrsim$ 10--100~m at 250--750~AU, the time scale to reach 
equilibrium is longer than the lifetime of the gaseous disk 
\citep[e.g.,][]{kb2015a}.  Here, we set $i_0$ = $e_0$/2 and adopt 
$e_0 = 10^{-4}$ at 250~AU and at 750~AU.  With these initial conditions, 
swarms of small particles are marginally gravitationally stable;
accretion begins in the shear regime. For the small Hill radii of 
100~km objects, however, initial accretion rates are fairly small.
\citet{kb2015a} demonstrate that modest changes to $e_0$ and $i_0$ 
have little impact on outcomes.

\subsection{Evolution at 250~AU}
\label{sec: evol-250}

In all calculations, the evolution follows a standard pattern. Large
oligarchs accrete pebbles fairly slowly. The small particles also merge 
and grow.  Along with the steady growth of an oligarch's `feeding zone' 
(set by the Hill radius), dynamical friction circularizes $e$ and $i$ 
for the oligarchs.  Gravitational focusing factors increase dramatically; 
runaway growth begins. During runaway growth, rapidly growing oligarchs 
stir up the more slowly growing pebbles. As gravitational focusing 
factors decline, the growth of oligarchs stalls. 

When runaway growth ends, oligarchs continue to accrete small particles.
Among the small particles, however, accretion and destructive collisions 
generate a broad size distribution with sizes ranging from 1~\mum\ to
1~km. As the evolution proceeds, collisional damping among 0.1--1.0~m 
and smaller particles overcomes gravitational stirring by the oligarchs
\citep[see also][]{kb2015a,kb2016}. Collisions among these small particles 
produce larger merged objects. Destructive collisions among the merged 
objects allow material to cycle between large and small objects, 
generating a roughly steady state size distribution. In addition to 
cycling of mass among 1~\mum\ to 1~km objects, destructive collisions
yield small objects ($r \lesssim$~1~\mum) which are ejected from the
system. Thus, the mass of the system declines with time.

Eventually, collisional damping of small particles leads to a second phase 
of runaway growth for the oligarchs. During this epoch, a few oligarchs 
grow from 2000--4000~km to $\gtrsim 10^4$~km. With more mass in oligarchs 
and less mass in small particles, gravitational stirring overcomes collisional
damping. Destructive collisions among all the small particles then power 
a collisional cascade, where the small particles are ground to dust and
ejected from the system. The growth of oligarchs ceases.

\begin{figure}
\includegraphics[width=6.5in]{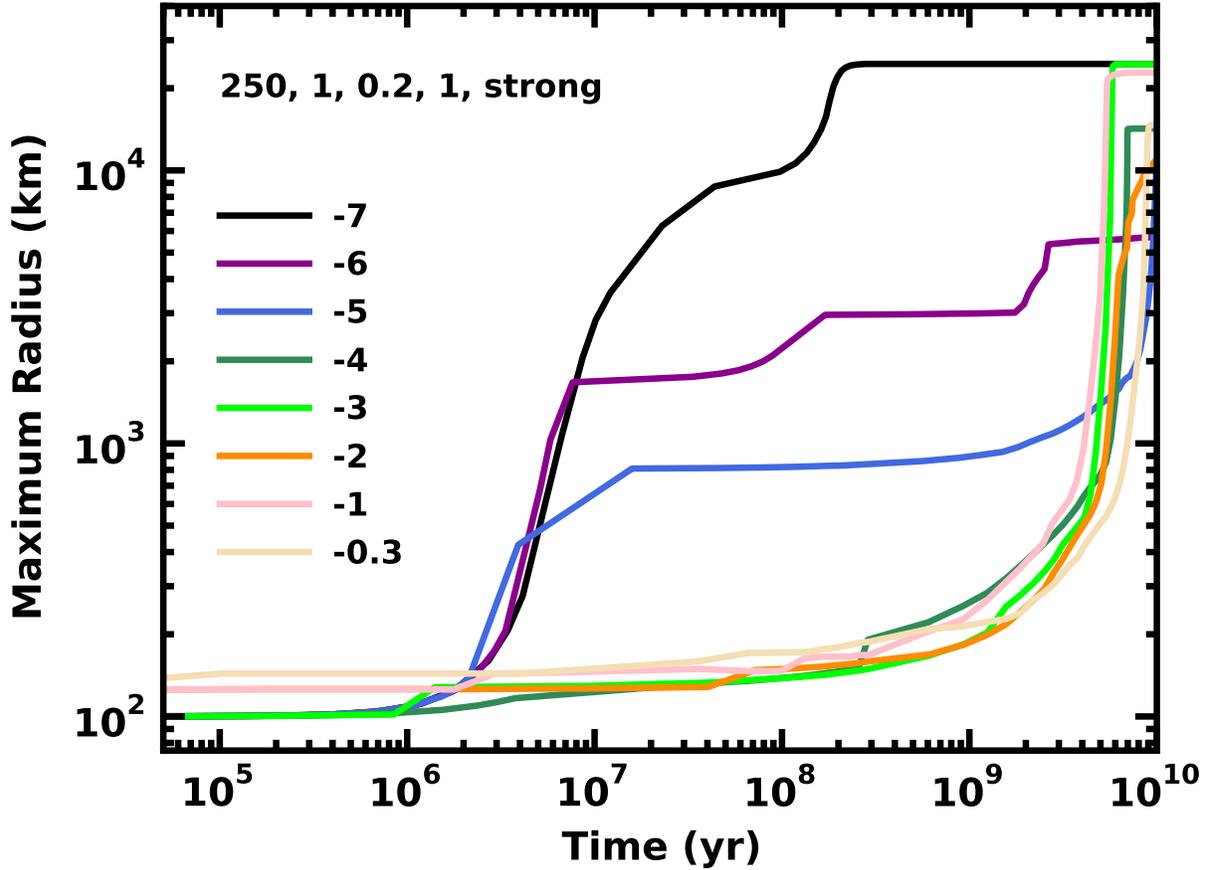}
\vskip 3ex
\caption{%
Growth of the largest object at 250~AU as a function of \f0, 
the initial mass fraction of solid material in 100~km objects, 
for calculations with \bd\ = 1, \ml0\ = 0.2, \bl\ = 1, and 
the strong fragmentation parameters. The legend indicates 
log~\f0\ for each calculation. When $ \f0\ \lesssim 10^{-5}$,
large objects grow rapidly; sometimes, these objects reach 
super-Earth masses with $\rmax\ \gtrsim 10^4$~km on short 
time scales. When $\f0\ \gtrsim 10^{-4}$, the largest objects 
grow slowly to super-Earth masses on time scales of 5--10~Gyr.
\label{fig: rmax1}
}
\end{figure}

Fig.~\ref{fig: rmax1} illustrates the growth of the largest objects in a
suite of calculations with strong planetesimals and various \f0. Among the
oligarchs, initial accretion rates are small, $\sim 10^{15}$~\gyr.  When 
$\f0\ \lesssim 10^{-5}$, steady growth and circularization lead to a first
runaway growth phase at 4~Myr. By 10--30~Myr, oligarchs have grown by 
factors of 10--100. Maximum sizes are much larger in systems starting with 
a few oligarchs than in those starting with many oligarchs.  From 50~Myr 
to a few Gyr, the largest oligarchs then grow slowly. When collisional 
damping allows a second phase of runaway growth, oligarchs reach maximum 
sizes of 3000~km to $2 \times 10^4$~km.

In systems with a larger fraction of the mass in oligarchs, growth is much
slower. When $\f0 \gtrsim 10^{-4}$, oligarchs rapidly stir themselves and 
the small pebbles. Gravitational focusing factors remain small, eliminating
the initial phase of runaway growth. As stirring continues, more and more
material is ejected or ends up in pebbles where collisional damping can 
overcome stirring by the oligarchs. Once damping overcomes stirring, large 
gravitational focusing factors enable a strong (but very late) phase of 
runaway growth where oligarchs grow from 200--300~km to 
$3 - 10 \times 10^3$~km sizes. Substantial mass loss prior to runaway growth
often limits the maximum sizes of oligarchs.

Despite the remarkable evolution during the second phase of runaway growth,
few systems produce super-Earth mass planets on time scales shorter than
the age of the solar system. Nearly all ensembles with 1--2 oligarchs yield
super-Earths in 200~Myr to 1~Gyr. Once the initial number of oligarchs 
exceeds 10, the second runaway growth phase is either too weak or too late
to enable super-Earth formation in $\lesssim$ 1--2~Gyr. 

\begin{figure}
\includegraphics[width=6.5in]{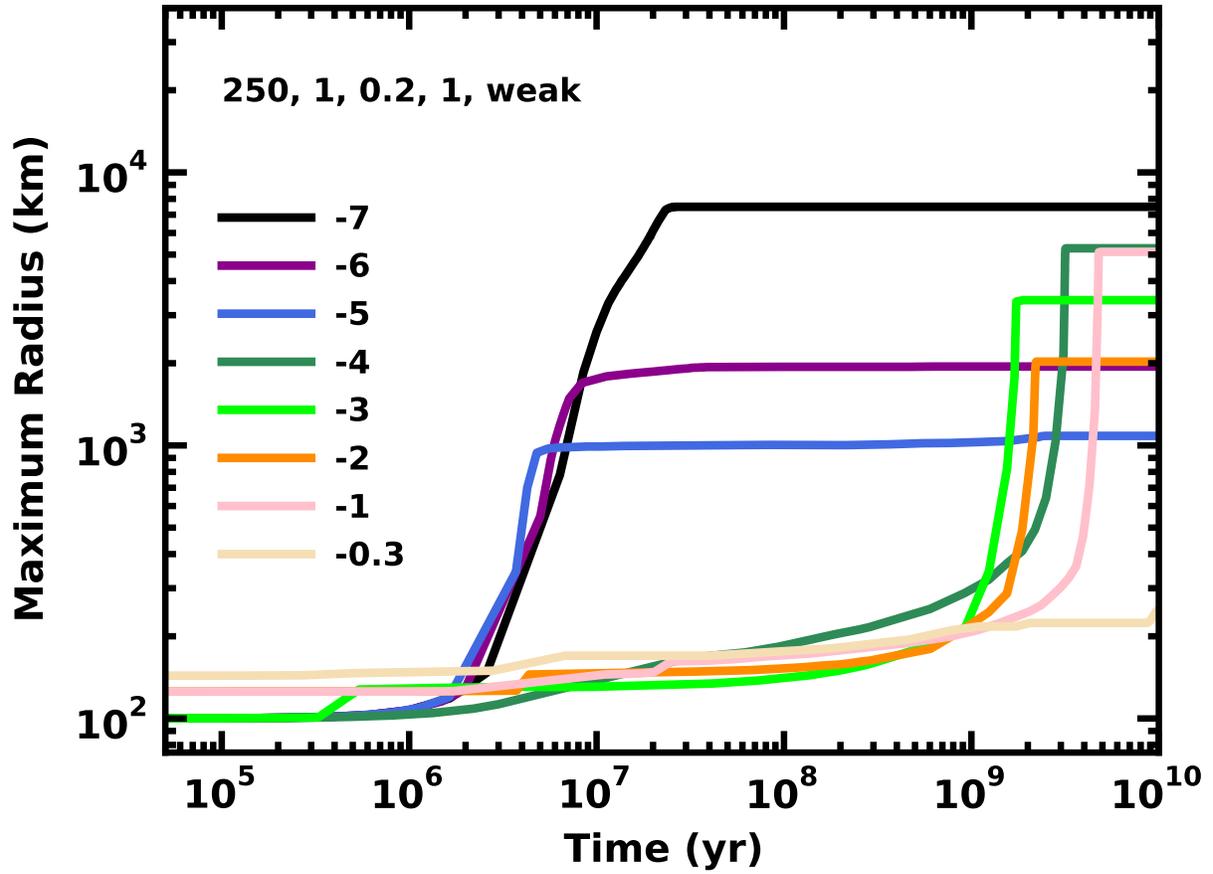}
\vskip 3ex
\caption{%
As in Fig. \ref{fig: rmax1} for calculations with the weak
fragmentation parameters. When the initial \f0\ is small 
(large), growth yields larger (smaller) planets. However, 
these systems rarely produce super-Earth mass planets.
\label{fig: rmax2}
}
\end{figure}

When planetesimals are much weaker, growth is less dramatic in systems 
with $\f0 \lesssim 10^{-5}$ (Fig.~\ref{fig: rmax2}). As in calculations 
with strong planetesimals, dynamical friction between oligarchs and pebbles 
initiates runaway growth of oligarchs at 3--5~Myr. After another 5--15~Myr, 
gravitational stirring by growing oligarchs raises the $e$ and $i$ of the 
pebbles and dramatically reduces gravitational focusing factors. The growth 
of oligarchs stalls.  When small planetesimals are weak and easy to break, 
stirring initiates the collisional cascade at an earlier epoch than in 
systems with strong planetesimals \citep[see also][]{kb2010}. Destructive 
collisions then result in greater mass loss from the system. Collisional
damping never overcomes gravitational stirring, preventing a second phase 
of runaway growth.  Compared to the most massive oligarchs in systems of 
strong planetesimals, the most massive oligarchs in these calculations are 
20--50 times less massive.

Systems with larger \f0\ evolve fairly independently of the planetesimal 
strength. In these systems, stirring by oligarchs maintains fairly large
$e$ and $i$ for pebbles and other small objects.  Larger $e$ and $i$ 
generates more debris and more mass loss.  All of these systems skip the 
first runaway growth phase at 3--10~Myr. As the evolution proceeds, mass
loss continues. In most systems, though, collisional damping slowly raises 
gravitational focusing factors. At 1--2~Gyr, there is a short period of 
runaway growth which produces 1--2 objects with radii of 1000--5000~km. 
Continued stirring initiates a stronger collisional cascade which grinds
the few remaining small objects to dust.

\begin{figure}
\includegraphics[width=6.5in]{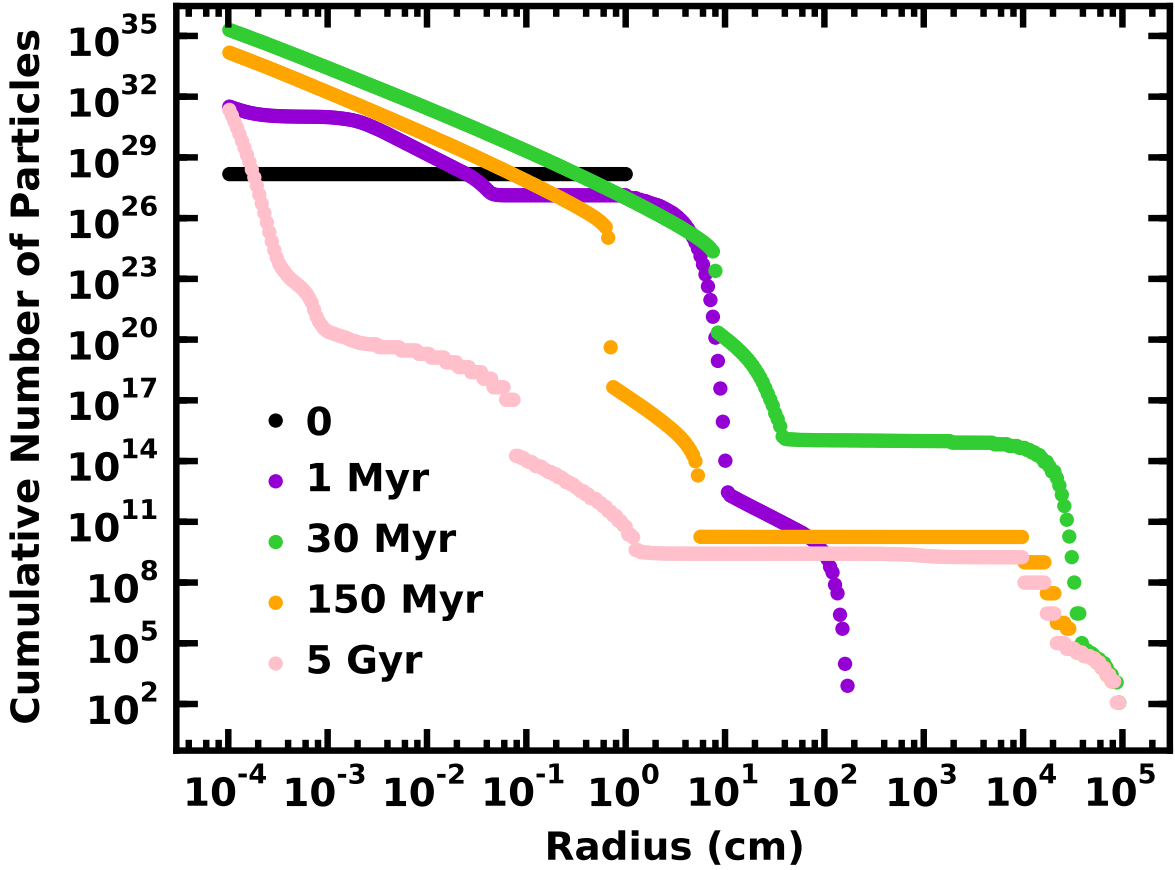}
\vskip 3ex
\caption{%
Time evolution of the cumulative size distribution for 
small particles with $r \lesssim$ 1~km at 250~AU in a 
simulation with $\f0\ = 10^{-7}$,
\bd\ = 1, \ml0\ = 0.2, \bl\ = 1, and strong planetesimals.
\label{fig: sd1}
}
\end{figure}

In all simulations of solid evolution at 250~AU, the evolution of small 
particles also follows a standard path (Fig.~\ref{fig: sd1}). Initially,
all pebbles have radii of 1~cm. After $\sim$ 1~Myr, growth and debris
production produce a multi-component size distribution extending from
\rmin\ = 1~\mum\ to roughly 1~m.  When the `pebbles' reach sizes of 1~km, 
stirring by oligarchs begins to drive a collisional cascade. Destructive
collisions first eliminate the weakest planetesimals with radii of $\sim$
0.1~km. As the cascade proceeds, stirring continues to raise collision
energies of large planetesimals; thus, collisions destroy progressively 
stronger (smaller) planetesimals. By 30~Myr, nearly all particles with 
radii of 1--100~m have been destroyed, producing a striking plateau in 
the cumulative size distribution. Debris from this population generates 
a power-law size distribution for pebbles with $r \lesssim$ 10~cm.

Until the second phase of runaway growth begins, destructive collisions 
and damping maintain two prominent features of the size distribution.  Among 
the smallest particles, the debris follows a power-law size distribution. At
the large end of this distribution, there is an abrupt drop in the cumulative 
number of particles. The particle radius at this drop, $r_d$, separates 
regions where collisional damping ($r \lesssim r_d$) and gravitational 
stirring ($r \gtrsim r_d$) dominate the velocity evolution. At larger sizes, 
destructive collisions produce debris; at smaller sizes, collisions produce 
larger merged objects. Solid material cycles between the two regions. 

\begin{figure}
\includegraphics[width=6.5in]{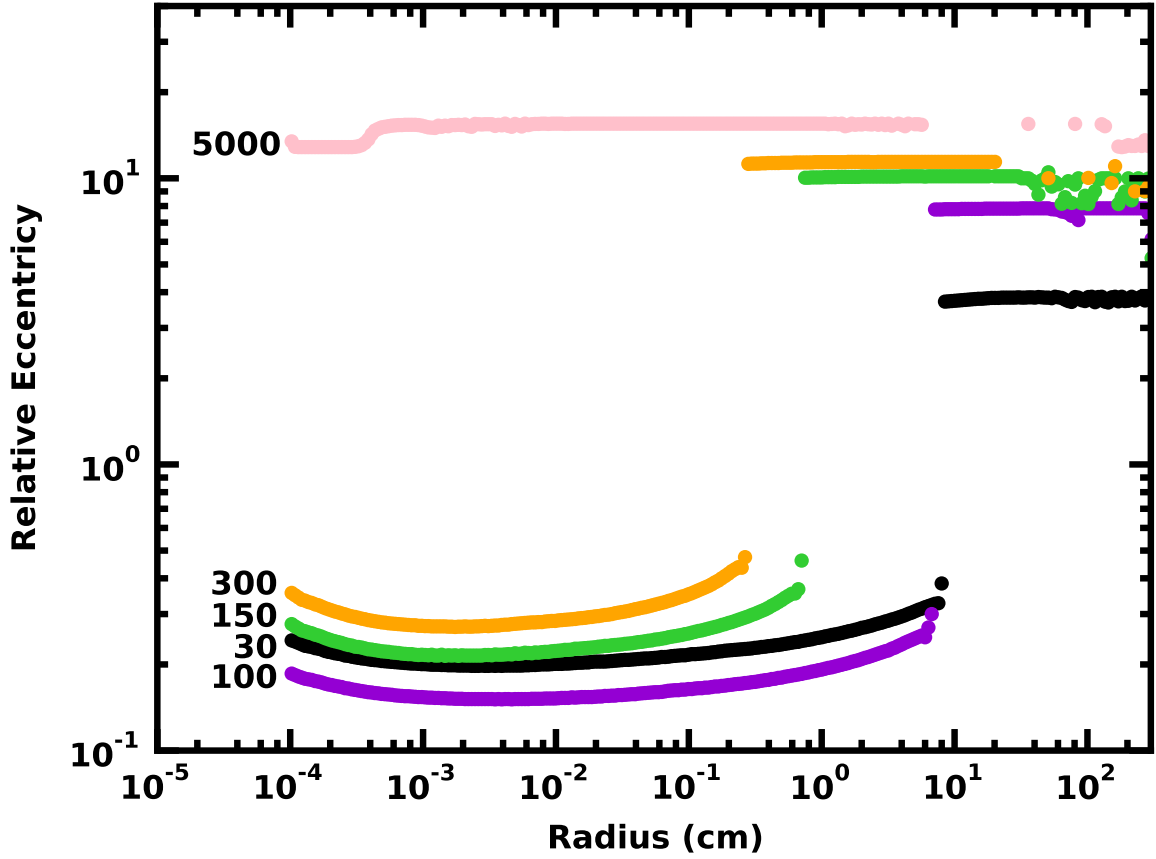}
\vskip 3ex
\caption{%
As in Fig.~\ref{fig: sd1} for the time evolution of the 
relative eccentricity distribution ($e_{rel} = e a / \rhill$, 
where \rhill\ is the Hill radius of the largest oligarch) 
for small particles with $r \lesssim$ 1~m at 250~AU.  Numbers 
to the left of each track indicate the evolution time in Myr.
\label{fig: vd1}
}
\end{figure}

Fig.~\ref{fig: vd1} illustrates the evolution of the `damping front' in
more detail \citep[see also][]{kb2015a}. At large sizes ($r \gtrsim r_d$),
gravitational stirring by massive oligarchs drives particle eccentricities 
to larger and larger values. In this example, the relative eccentricity
grows from $e_{rel} \approx$ 4 to $e_{rel} \approx$ 15 as the mass of the
most massive oligarch grows by a factor of 50. Among smaller particles, 
collisional damping produces a sharp, factor of 10--20, drop in the relative 
eccentricity. As the system evolves, the damping front moves to smaller 
and smaller particle radii.  Once these particles contain enough mass, they 
power a second phase of runaway growth where the most massive oligarchs 
reach super-Earth masses. Eventually, the mass in small particles vanishes.
Gravitational stirring dominates collisional damping; the damping front 
disappears. All small particles then have roughly the same $e_{rel}$.

\begin{figure}
\includegraphics[width=6.5in]{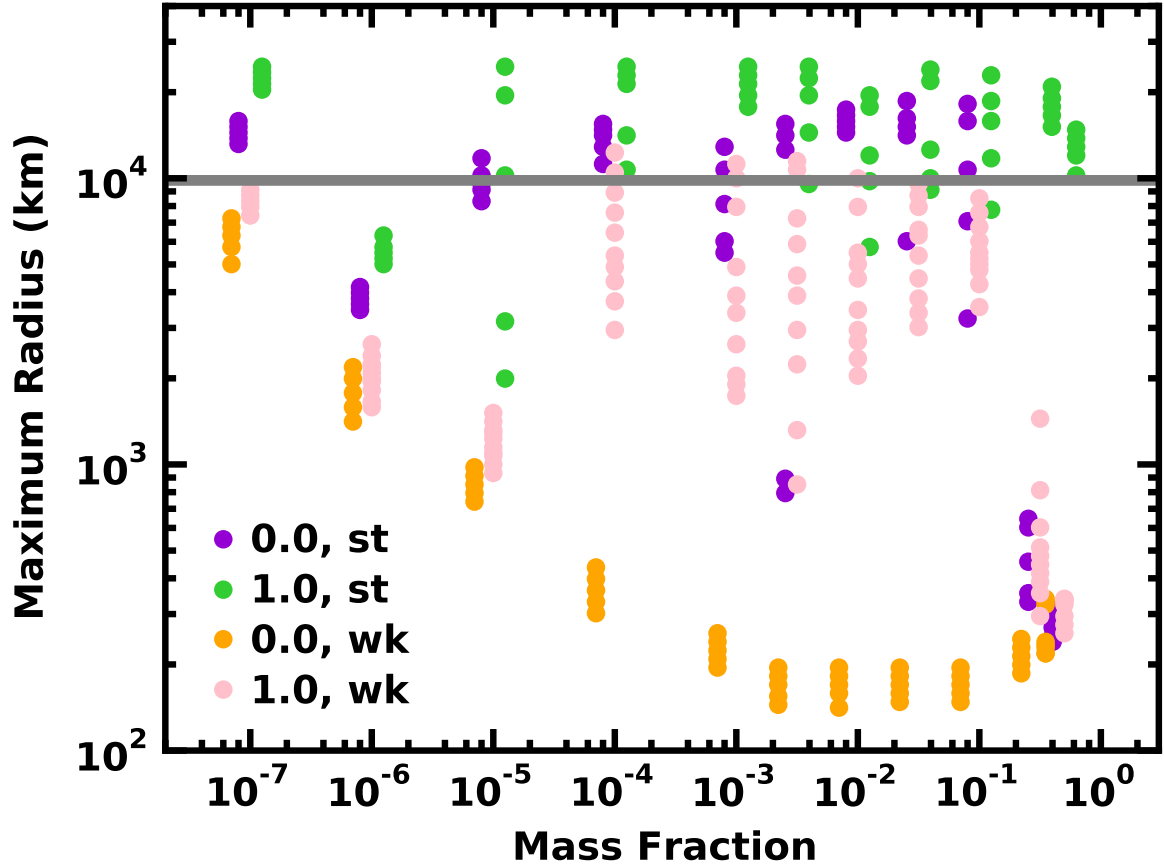}
\vskip 3ex
\caption{%
Maximum radius \rmax\ as a function of the mass fraction
\f0, \bl, and the strength of small planetesimals for 
calculations at 250~AU. The legend indicates \bl\ and 
the planetesimal strength (`st' for strong; `wk' for weak).  
Super-Earth formation generally requires strong planetesimals.
\label{fig: rmax-all}
}
\end{figure}

To summarize results for the complete suite of simulations, 
Fig.~\ref{fig: rmax-all} shows \rmax\ at the end of each calculation 
as a function of \f0, \bl, and
the planetesimal strength. Among calculations with weak planetesimals
and \bl\ = 0, there is a clear trend of less growth with more initial 
mass in large oligarchs. In these calculations, more oligarchs produce
more gravitational stirring among the pebbles. With more gravitational
stirring, pebbles are easier to break as they grow from 1--10~cm to 
10~m to 1~km sizes. Collisions then produce more debris, robbing the
massive oligarchs of material to grow to super-Earth sizes. 

When \bl\ = 1 and planetesimals are weak, some calculations follow the
trend established for systems with \bl\ = 0. For simulations with \f0 = 
$1-100 \times 10^{-7}$ and \f0\ $\approx$ 0.3--0.5, gravitational 
stirring by oligarchs leads to destructive collisions which remove 
significant amounts of mass from the annulus. Oligarch growth then 
stalls at small sizes. Among the simulations with intermediate \f0,
collisional damping associated with copious amounts of debris often 
overcomes gravitational stirring, enabling the system to retain small
particles and energizing a second phase of runaway growth and the
production of Earth mass planets. In these simulations, there is a
broad range of outcomes: Pluto to Mars mass planets are just as likely
as Earth mass planets.

To understand the long-term evolution of systems of weak planetesimals
in more detail,
we performed an additional set of calculations with $Q_b$ = 10~\ergg\ and
\bl\ = 1. In these systems, planetesimals begin to break at even earlier
times than those with $Q_b$ = $10^3$~\ergg. Mass loss is more severe;
collisional damping never overcomes gravitational stirring. The variation
of \mmax\ with \f0\ then almost precisely follows results for calculations
with weak planetesimals and \bl\ = $0$ in Fig.~\ref{fig: rmax-all},
where no oligarchs reach super-Earth masses.

Within the suite of calculations with strong planetesimals, super-Earth
formation is common. Nearly all simulations with \bl\ = 1 yield a 
super-Earth. Although most simulations with \bl\ = 0 produce a 
super-Earth, the largest objects in calculations with \f0\ = 0.3--0.5 
always remain small.
Despite the overall success of these simulations, rapid super-Earth
production is still rare. In all calculations with \f0\ $\gtrsim 10^{-6}$,
protoplanets reach super-Earth masses on 5--10~Gyr time scales, longer 
than the age of the solar system. When systems begin with $\lesssim$ 
5--10 massive oligarchs, a single massive protoplanet can grow to 
super-Earth masses in 100--300~Myr. 

\subsection{Evolution at 750~AU}
\label{sec: evol-750}

\begin{figure}
\includegraphics[width=6.5in]{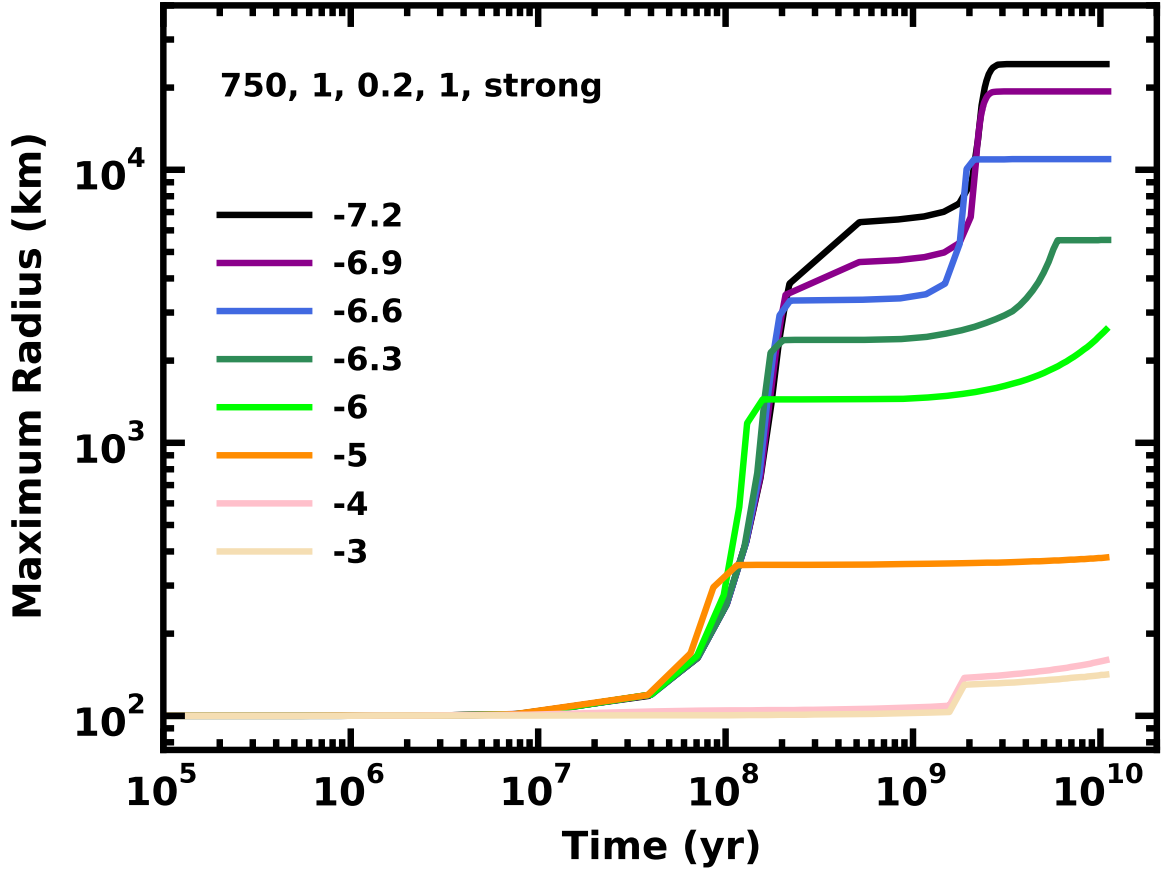}
\vskip 3ex
\caption{%
Growth of the largest object at 750~AU as a function of \f0, 
the initial mass fraction of solid material in 100~km objects, 
for calculations with \bd\ = 1, \ml0\ = 0.2, \bl\ = 1, and the 
strong fragmentation parameters. The legend indicates log~\f0.
Simulations with $ \f0\ \lesssim 10^{-5}$ produce 300--4000~km 
objects in 100--200~Myr. After a 1--5~Gyr period where the 
largest objects grow very slowly, simulations with 
$\f0\ \lesssim 5 \times 10^{-7}$ undergo a second phase of 
runaway growth. The largest objects may then reach super-Earth 
sizes with $\rmax\ \gtrsim 10^4$~km.
\label{fig: rmax3}
}
\end{figure}

Within an annulus at 750~AU, oligarchs follow the same evolutionary path 
as at 250~AU (Fig.~\ref{fig: rmax3}).  At the start of the calculations,
100~km oligarchs accrete material in the shear regime. With typical growth 
rates of $2 - 6 \times 10^{13}$~g~yr$^{-1}$, it takes 100--600~Myr for 
oligarchs to double in mass. As they grow, oligarchs try to stir up the 
much smaller pebbles. Collisional damping among the pebbles counters this
stirring.

In systems with a few oligarchs ($\f0 \lesssim 10^{-5}$) and strong pebbles, 
collisional damping dominates gravitational stirring. Once oligarchs have 
2--4 times their initial masses, runaway growth begins. Within a few Myr, 
oligarchs reach sizes of 400--4000~km. Throughout runaway growth, 
gravitational stirring rates also grow. Eventually, stirring overcomes 
damping. Runaway growth ends. 

As runaway growth ends, destructive collisions among leftover pebbles and
planetesimals drive a collisional cascade. When the number of oligarchs 
is small ($\lesssim$ 10--20), collisional damping among small particles
once again dominates 
gravitational stirring by the oligarchs. Damping powers a second phase of 
runaway growth, where oligarchs may reach super-Earth masses. The onset of 
this epoch depends on the number of oligarchs. Systems with 1--4 oligarchs
reach super-Earth masses in 1--2 Gyr. Although systems with 8 or more 
oligarchs enter a second phase of runaway growth, this evolution never 
leads to super-Earth mass planets.

In systems with many oligarchs, stirring dominates damping. Oligarch masses 
grow only by factors of 2--3 over 1--2~Gyr, runaway growth never develops.
After 10~Gyr, oligarchs reach sizes of 150--200~km. 

\begin{figure}
\includegraphics[width=6.5in]{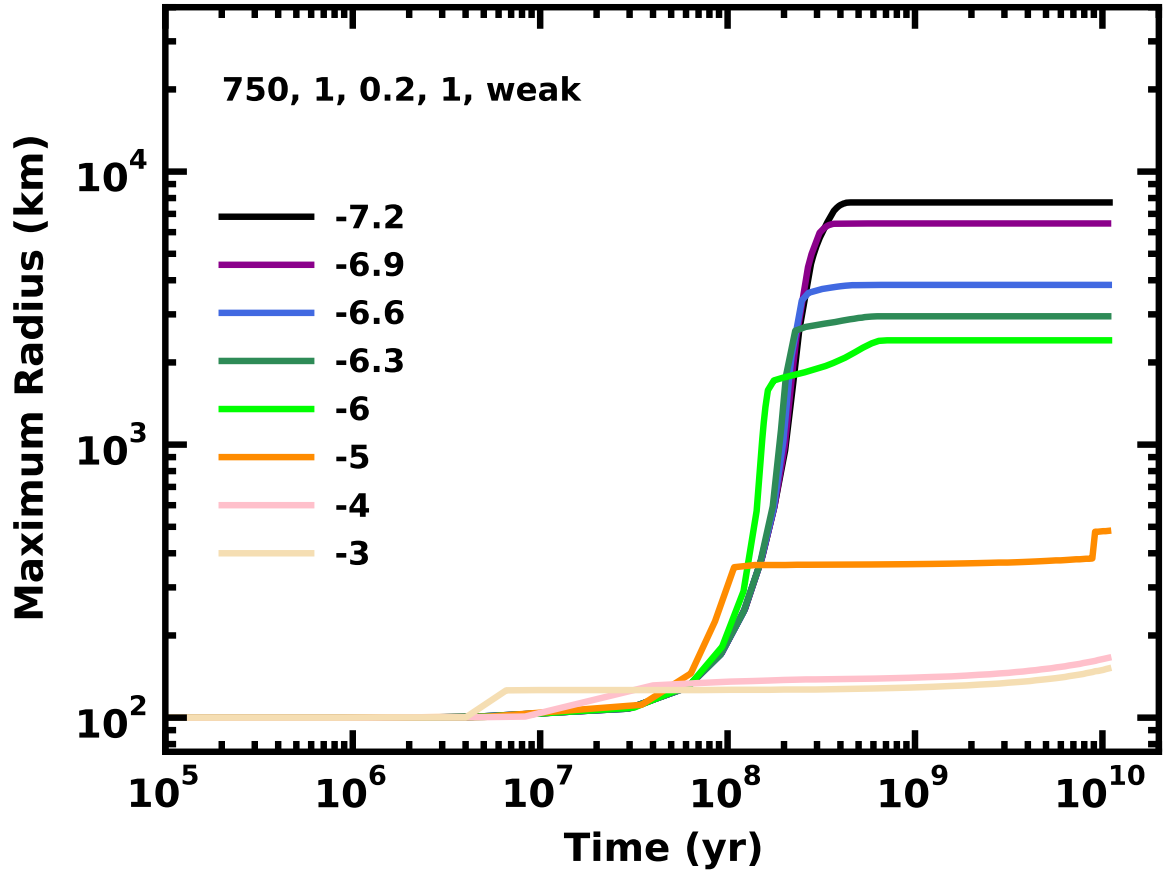}
\vskip 3ex
\caption{%
As in Fig.~\ref{fig: rmax3} for calculations with weak planetesimals.
Although \rmax\ correlates inversely with the initial mass in 
oligarchs, the largest objects never reach super-Earth masses.
\label{fig: rmax4}
}
\end{figure}

When pebbles at 750~AU are weak (Fig.~\ref{fig: rmax4}), the evolution never 
leads to super-Earth mass planets.  Although the initial phase of runaway growth
produces massive planets, the subsequent collisional cascade effectively
destroys leftover pebbles and more massive planetesimals. In these systems,
collisional damping is never effective enough to create the pronounced 
damping front observed in calculations with strong pebbles. Collisions 
destroy pebbles and smaller objects faster than oligarchs can accrete them.
Thus, growth stalls at much smaller masses.

For calculations with strong or weak planetesimals, the timing of the 
first phase of runaway growth always occurs 30--40 times later at 750~AU 
(120--150~Myr) than at 250~AU (4~Myr).  When \r0, \rmax, \f0, $e_0$, 
and $i_0$ are identical, the time for runaway growth to produce objects
with a fixed size depends on the orbital period, surface density, and
the gravitational focusing factor $t \propto  P / f_g \Sigma$ 
\citep[e.g.,] []{liss1987,kb2008}. Given our starting conditions, initial
gravitational focusing factors at 250~AU and at 750~AU are roughly equal;
$P / \Sigma$ is roughly 45 times larger at 750~AU than at 250~AU.  Despite 
stochastic variations in the collision, damping, and stirring rates, the 
predicted factor of 45 difference in the timing of runaway growth is 
remarkably close to the factor of 30--40 difference of our calculations.

\subsection{Luminosity Evolution} 

Aside from detecting planets by direct imaging \citep[e.g.,][]{marois2008,
lafren2010,lagrange2010,bailey2014,currie2014a,currie2014b}, 
scattered light and thermal emission from small particles are the only 
diagnostic of the long-term evolution of protoplanets at large $a$. 
In our calculations, we track the size distribution of 1~\mum\ and 
larger particles.  Relative to the luminosity of the central star, 
the dust luminosity is 
\begin{equation}
L_d / \lstar = \ad\ / 4 \pi a^2 ~,
\label{eq: ldust}
\end{equation}
where \ad\ is the cross-sectional area of the swarm of particles. For a
dust albedo $w$, the luminosity in scattered light is 
$w \ldlstar$; the thermal emission is $(1 - w) \ldlstar$.

\begin{figure}
\includegraphics[width=6.5in]{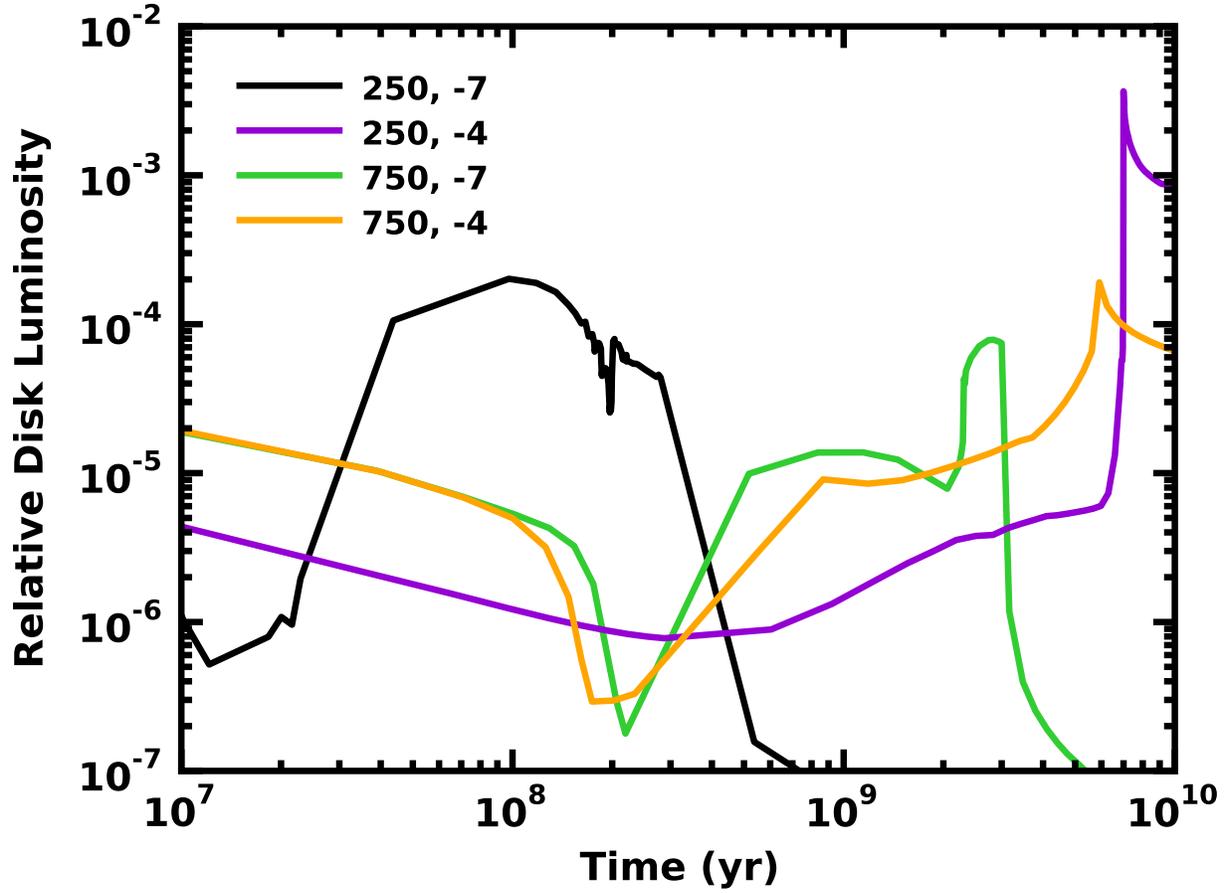}
\vskip 3ex
\caption{%
Evolution of the relative dust luminosity \ldlstar\ for 
calculations at 250--750~AU. The legend indicates $a$ 
and log~\f0\ for each model.  Typical maximum dust 
luminosities range from 
$\ldlstar\ \approx 10^{-5} - 10^{-3}$ at 100~Myr to 
$\ldlstar\ \approx 10^{-9} - 10^{-3}$ at 10~Gyr. 
\label{fig: ldust}
}
\end{figure}

Fig.~\ref{fig: ldust} illustrates the long-term evolution of \ldlstar\ for
several calculations. At 250~AU, rings of pebbles with \f0\ = $10^{-7}$ have
an initial $\ldlstar\ \approx 1-3 \times 10^{-4}$ (Fig.~\ref{fig: ldust},
black curve). As the system evolves, small particles merge into larger 
objects. The dust luminosity gradually drops, reaching 
$ \ldlstar\ \approx 3 - 10 \times 10^{-7}$ after roughly 10~Myr. Near the 
end of the first epoch of runaway growth, the dust brightens by more than 
two orders of magnitude. After maintaining a peak 
$\ldlstar \approx 1 - 3 \times 10^{-4}$ for 50--100~Myr, the system starts
to fade. During the decline, the second runaway growth phase results in a 
small short-lived brightening of the dust luminosity. Once runaway growth 
ends, the dust fades rapidly to very faint levels with $\ldlstar \lesssim$
$10^{-7}$.

In systems with more oligarchs (larger \f0), peak dust luminosity 
occurs later and later in time. For \f0\ = $10^{-4}$ 
(Fig.~\ref{fig: ldust}, violet curve), slower growth 
of smaller particles results in a more gradual fading of the dust 
luminosity. After reaching a minimum $\ldlstar \approx 10^{-6}$ at 
300--500~Myr, continued stirring by slowly growing oligarchs creates a
more vigorous collisional cascade and a slowly rising \ldlstar. After
5--6~Gyr, a last gasp of runaway growth powers a steeply rising `burst' in 
the dust production rate where \ldlstar\ rises to $1 - 4 \times 10^{-3}$.
As the collisional cascade continues, \ldlstar\ drops.

At 750~AU, the evolution is slower and more muted. In all systems of
pebbles, the dust luminosity slowly declines from an initial value of
$\ldlstar \approx 3 \times 10^{-5}$ to $1 - 3 \times 10^{-7}$ at 
200--300~Myr. Once runaway growth stalls, gravitational stirring by 
the oligarchs increases the velocities of leftover pebbles and 
planetesimals. Destructive collisions generate copious amounts of
small particles; \ldlstar\ rises by an order of magnitude and then 
remains roughly constant. During the second epoch of runaway growth, 
the dust emission rises by another order of magnitude, reaching 
$\ldlstar \approx 10^{-4}$ at 2--3~Gyr 
(\f0\ = $10^{-7}$; Fig.~\ref{fig: ldust}, green curve) to
6~Gyr (\f0\ = $10^{-4}$; Fig.~\ref{fig: ldust}, orange curve) to
10~Gyr (\f0\ = $10^{-3}$). Following runaway growth, resumption 
of the collisional cascade leads to a rapid drop in the dust luminosity.

\section{DISCUSSION}
\label{sec: disc}

Together with \citet{bk2014}, \citet{kb2015a}, and \citet{bk2016}, 
we have examined several plausible mechanisms which yield a 
super-Earth mass planet at 100--750~AU around a solar-type star. In 
{\it scattering} scenarios, multiple super-Earths form at 3--20~AU 
\citep[e.g.,][]{bk2011a}. As a few of these accrete gas and grow 
into gas giants, they scatter lower mass protoplanets into high $e$ 
orbits \citep[see also][and references therein]{rasio1996,weiden1996,
chatterjee2008,ford2008,moeckel2008,marzari2010,naga2011,moeckel2012}.
Interactions with the gaseous disk circularize the scattered
protoplanet's orbit at large $a$. The {\it in situ} models begin 
with a ring of solid material at $a$ = 100--750~AU.  Collisional 
growth produces super-Earth mass planets on time scales which 
depend on the semimajor axis and initial mass of the ring and the 
initial sizes of the solids. 

Both approaches successfully produce super-Earths with $a \approx$ 
100--500~AU. Scattering allows super-Earths to reach large $a$ on
short time scales, $\lesssim$ 10~Myr. However, orbits are often 
eccentric, $e \gtrsim$ 0.1. Although {\it in situ} growth enables 
more circular orbits, growth times range from $\sim$ 100~Myr to 
$\gtrsim$ 10~Gyr.

\subsection{Other Approaches to Super-Earth Formation}

Among alternatives to these scenarios, gravitational instability 
in a massive circumstellar disk is probably the most popular
\citep[e.g.,][and references therein]{helled2014,rice2016}. 
Despite considerable effort to understand the formation and 
evolution of clumps in an unstable disk, relating specific 
outcomes to initial conditions remains uncertain. For 
super-Earth mass planets beyond 100~AU, tidally downsizing a 
Jupiter mass clump is essential \citep{nayak2010,nayak2015}. 
Achieving this goal is also uncertain \citep{forgan2013,nayak2015}.

Although not directly designed to address the formation of planet 
nine, other scattering mechanisms appear capable of placing a massive
planet on a high $e$ orbit at $a \approx$ 250--1000~AU.  Current 
models for the Oort cloud rely on scattering of leftover 1~km and 
larger objects from orbits near the original locations of the gas 
giants \citep[e.g.,][]{ida2000a,morbi2004,brass2006,levison2010b,brass2012}.  
Dynamical interactions with nearby stars then stabilize comets 
within the Oort cloud. Numerical simulations of these processes 
often yield objects with orbits similar to Sedna, \VP, and other 
solar system objects on high $e$ orbits with $a \gtrsim$ 150--200~AU.

Other investigations consider the possibility of capturing Sedna, 
\VP, and planet nine during a stellar flyby \citep[e.g.,][]{morbi2004,
kb2004d,jilkova2015,li2016}. Although numerical simulations often 
yield planets and dwarf planets with reasonable orbits, the trajectory
and distance of closest approach for the passing star must be tuned 
to achieve these orbits. Some encounters also leave behind captured 
and indigenous objects on orbits which are inconsistent with the
current inventory of trans-Neptunian objects.

All planet nine theories must consider the long-term evolution of 
orbits in the inner and outer solar system. Single or multiple 
super-Earth mass planets at $a \approx$ 100--300~AU are probably 
inconsistent with the ephemerides of Jupiter, Saturn, and other major
planets \citep[e.g.,][]{iorio2012,delafm2014,iorio2014,fienga2016}.
However, certain ranges for the true anomaly of a single super-Earth 
with $a \approx$ 500~AU {\it improve} the Cassini residuals for Saturn
\citep{fienga2016}. If planet nine is ever detected, Cassini radio
ranging data will provide a strong constraint on the allowed mass
and orbital parameters.

In the outer solar system, close encounters with nearby stars can
perturb the orbital elements of planets with $a \gtrsim$ 200~AU
\citep[e.g.,][]{morbi2004,kb2004d,brass2012,jilkova2015,li2016}. 
Using an extensive set of numerical simulations, \citet{li2016} 
conclude that interactions with passing stars might strip planet 
nine from the solar system.  However, they do not address how 
these interactions impact the orbits of Sedna, \VP, and comets 
within the Oort cloud. Thus, it is not clear whether typical
outcomes of these simulations are consistent with observations
of solar system objects.

Additional theoretical investigations are clearly needed to 
examine the history of the outer solar system in the context 
of current observations. Improvements in our understanding 
of disk dissipation can help us explore the evolution of gas 
at 100--1000~AU on time scales when collisional growth or 
scattering might place super-Earths in the outer disk.  As 
disks dissipate, more detailed studies of the expansion of 
the inner cavity provide a way to relate the orbits of small 
solids and planets to the evolution of the disk. Finally,
broader studies of the outcomes of stellar encounters enable
a better understanding of the current architecture of the
outer solar system.

\subsection{Observational Tests: Debris Disks}

Although direct imaging techniques have discovered many gas giants,
they cannot
detect super-Earth mass planets. However, observations of structure 
in debris disks place some limits on the formation of super-Earths 
at $a \gtrsim$ 100~AU around solar-type stars \citep[][and 
references therein]{kb2015a}.  Constraints on the observed dust 
luminosity also provide tests of our numerical calculations.

In \citet{kb2015a}, we focused on three solar-type stars -- HD~107146, 
HD~202628, and HD~207129 -- with large rings of debris at 100--200~AU
\citep{corder2009,krist2010,marshall2011,krist2012,ricci2015}. With
ages of 1--2~Gyr and relative dust luminosities 
$\ldlstar \approx 10^{-4}$, HD~202628 and HD~207129 provide 
interesting comparisons with our model predictions. Although 
the maximum dust luminosity in calculations with a single 
oligarch match the observations, these systems achieve peak 
\ldlstar\ too early -- 100~Myr -- and fade too rapidly. Models
with 16 or more oligarchs match the observed \ldlstar\ when 
the central star is too old, $\gtrsim$ 5~Gyr.  However, models 
with 2--8 oligarchs reach the observed \ldlstar\ at 1--2~Gyr 
and remain bright for several Gyr. Thus, these models yield 
a reasonable match to observations.

Several of our calculations match the observed dust luminosity 
for the debris disk in the 100~Myr old star HD~107146. With
$\ldlstar \approx 10^{-3}$ \citep{will2004}, this system is
among the brightest debris disks around a solar-type star. 
Aside from our `standard' debris disk models starting from
ensembles of 1~km planetesimals embedded in a gaseous disk
\citep{kb2008,kb2010}, calculations of several oligarchs 
within a ring of pebbles at 100--150~AU yield 
$\ldlstar\ \approx 10^{-3}$ at roughly 100~Myr.

Observations with ALMA will certainly improve these tests. As
samples of solar-type stars with resolved debris disks beyond
100~AU grow, high spatial resolution observations should yield
better comparisons with predictions of the surface density
distribution. Robust estimates of the frequency and sizes of
the dark lanes produced by planets can also test theoretical
models.

\subsection{Observational Tests: Solar System}

Future observations will clarify the populations of planets beyond 
100~AU.  Current large-format optical imagers are capable of detecting 
planet nine and many other Sedna-like dwarf planets with an albedo of 
0.05--0.3.  For many of these objects, infrared detections with the 
{\it James Webb Space Telescope} should yield robust measurements of
the albedo and radius. Within 10--15 yr, data from the 
{\it Large Synoptic Survey Telescope} will provide much larger 
samples and test our understanding of the long-term dynamics of 
the outer solar system.

Direct detection of planet nine clearly tests scenarios for 
super-Earth formation beyond 100~AU. A Super-Earth on a nearly 
circular orbit favors {\it in situ} formation scenarios. 
Eccentric orbits favor scattering models. For any orbit, 
the ephemerides of Jupiter, Saturn, and other gas giants place
strong limits on the mass \citep[e.g.,][]{iorio2014,fienga2016}.

As our understanding of the dwarf planet population at 100--1000~AU
improves, comparisons of measured orbital parameters with results 
from long-term scattering simulations should provide tests of models
with different evolutionary histories. Dwarf planets at high 
inclination provide a particularly stringent test, placing 
constraints on the initial mass in solids at large $a$ and
the encounter history of the outer solar system
\citep{jilkova2015,madigan2016,batygin2016,li2016,brown2016}.

\section{SUMMARY}
\label{sec: summ}

We use a suite of coagulation calculations to isolate paths for 
{ \it in situ} production of 
super-Earth mass planets at 250--750~AU around solar-type stars. 
These paths begin with a massive ring, $M_0 \gtrsim$ 15~\mearth, 
composed of strong pebbles, $r_0 \approx$ 1~cm, and a few large 
oligarchs, $r \approx$ 100~km. When these systems contain 1--10 
oligarchs, two phases of runaway growth yield super-Earth mass 
planets in 100--200~Myr at 250~AU and 1--2~Gyr at 750~AU. Large 
numbers of oligarchs stir up the pebbles and initiate a collisional
cascade which prevents the growth of super-Earths. For any number 
of oligarchs, systems of weak pebbles are also incapable of 
producing a super-Earth mass planet in 10~Gyr.

The debris from swarms of pebbles producing super-Earths at 
250--750~AU is directly visible. These systems have relative 
dust luminosities $\ldlstar \approx 1-30 \times 10^{-4}$ at 
ages of 100~Myr to 10~Gyr.  Within the rings of dust generated 
by planet growth, super-Earths should create gaps in the 
surface density distribution. Predicted widths for the gaps 
are 10--20~AU at 250~AU and 30--60~AU at 750~AU.

Over the next decade, observations can test this scenario.  Among 
exoplanetary systems, discovering super-Earths, gas giants, or 
debris disks far from their host stars provide vital information 
on the long-term evolution of protoplanets and circumstellar disks.
In the solar system, orbital parameters for newly discovered dwarf
planets with $a \gtrsim$ 100~AU allow more rigorous tests of 
proposals for planet nine \citep[e.g.,][]{batygin2016,brown2016}. 
If planet nine is real, direct detection constrains models for 
{\it in situ} formation and scattering 
\citep[see also][]{bk2014,kb2015a,li2016,bk2016}. 

\vskip 6ex

We acknowledge generous allotments of computer time on the NASA `discover' 
cluster.  Advice and comments from M. Geller, J. Najita, and D. Wilner 
greatly improved our presentation.  Portions of this project were supported 
by {\it NASA Outer Planets Program} through grant NNX11AM37G.

\bibliography{ms.bbl}

\begin{thebibliography}{}
\expandafter\ifx\csname natexlab\endcsname\relax\def\natexlab#1{#1}\fi

\bibitem[{{Adachi} {et~al.}(1976){Adachi}, {Hayashi}, \& {Nakazawa}}]{ada76}
{Adachi}, I., {Hayashi}, C., \& {Nakazawa}, K. 1976, Progress of Theoretical
  Physics, 56, 1756

\bibitem[{{Alexander} {et~al.}(2014){Alexander}, {Pascucci}, {Andrews},
  {Armitage}, \& {Cieza}}]{alex2014}
{Alexander}, R., {Pascucci}, I., {Andrews}, S., {Armitage}, P., \& {Cieza}, L.
  2014, in Protostars and Planets VI, ed. H.~{Beuther}, R.~S. {Klessen}, C.~P.
  {Dullemond}, \& T.~{Henning} (University of Arizona Press, Tucson, AZ),
  475--496

\bibitem[{{Andrews}(2015)}]{andrews2015}
{Andrews}, S.~M. 2015, \pasp, 127, 961

\bibitem[{{Andrews} {et~al.}(2011){Andrews}, {Wilner}, {Espaillat}, {Hughes},
  {Dullemond}, {McClure}, {Qi}, \& {Brown}}]{andrews2011}
{Andrews}, S.~M., {Wilner}, D.~J., {Espaillat}, C., {et~al.} 2011, \apj, 732,
  42

\bibitem[{{Arakawa} {et~al.}(2002){Arakawa}, {Leliwa-Kopystynski}, \&
  {Maeno}}]{arakawa2002}
{Arakawa}, M., {Leliwa-Kopystynski}, J., \& {Maeno}, N. 2002, \icarus, 158, 516

\bibitem[{{Ayliffe} {et~al.}(2012){Ayliffe}, {Laibe}, {Price}, \&
  {Bate}}]{ayliffe2012}
{Ayliffe}, B.~A., {Laibe}, G., {Price}, D.~J., \& {Bate}, M.~R. 2012, \mnras,
  423, 1450

\bibitem[{{Bailey} {et~al.}(2014){Bailey}, {Meshkat}, {Reiter}, {Morzinski},
  {Males}, {Su}, {Hinz}, {Kenworthy}, {Stark}, {Mamajek}, {Briguglio}, {Close},
  {Follette}, {Puglisi}, {Rodigas}, {Weinberger}, \& {Xompero}}]{bailey2014}
{Bailey}, V., {Meshkat}, T., {Reiter}, M., {et~al.} 2014, \apjl, 780, L4

\bibitem[{{Batygin} \& {Brown}(2016)}]{batygin2016}
{Batygin}, K., \& {Brown}, M.~E. 2016, \aj, 151, 22

\bibitem[{{Benz} \& {Asphaug}(1999)}]{benz1999}
{Benz}, W., \& {Asphaug}, E. 1999, Icarus, 142, 5

\bibitem[{{Birnstiel} \& {Andrews}(2014)}]{birnstiel2014}
{Birnstiel}, T., \& {Andrews}, S.~M. 2014, \apj, 780, 153

\bibitem[{{Birnstiel} {et~al.}(2010){Birnstiel}, {Dullemond}, \&
  {Brauer}}]{birn2010}
{Birnstiel}, T., {Dullemond}, C.~P., \& {Brauer}, F. 2010, \aap, 513, A79+

\bibitem[{{Bottke} {et~al.}(2010){Bottke}, {Nesvorn{\'y}}, {Vokrouhlick{\'y}},
  \& {Morbidelli}}]{bottke2010}
{Bottke}, W.~F., {Nesvorn{\'y}}, D., {Vokrouhlick{\'y}}, D., \& {Morbidelli},
  A. 2010, \aj, 139, 994

\bibitem[{{Brasser} {et~al.}(2006){Brasser}, {Duncan}, \&
  {Levison}}]{brass2006}
{Brasser}, R., {Duncan}, M.~J., \& {Levison}, H.~F. 2006, \icarus, 184, 59

\bibitem[{{Brasser} {et~al.}(2012){Brasser}, {Duncan}, {Levison}, {Schwamb}, \&
  {Brown}}]{brass2012}
{Brasser}, R., {Duncan}, M.~J., {Levison}, H.~F., {Schwamb}, M.~E., \& {Brown},
  M.~E. 2012, \icarus, 217, 1

\bibitem[{{Brauer} {et~al.}(2008){Brauer}, {Dullemond}, \&
  {Henning}}]{brauer2008}
{Brauer}, F., {Dullemond}, C.~P., \& {Henning}, T. 2008, \aap, 480, 859

\bibitem[{{Bromley} \& {Kenyon}(2006)}]{bk2006}
{Bromley}, B.~C., \& {Kenyon}, S.~J. 2006, \aj, 131, 2737

\bibitem[{{Bromley} \& {Kenyon}(2011)}]{bk2011a}
---. 2011, \apj, 731, 101

\bibitem[{{Bromley} \& {Kenyon}(2014)}]{bk2014}
---. 2014, \apj, 796, 141

\bibitem[{{Bromley} \& {Kenyon}(2016)}]{bk2016}
---. 2016, \apj, submitted (available on the arXiv)

\bibitem[{{Brown} \& {Batygin}(2016)}]{brown2016}
{Brown}, M.~E., \& {Batygin}, K. 2016, arXiv:1603.05712

\bibitem[{{Brown} {et~al.}(2004){Brown}, {Trujillo}, \&
  {Rabinowitz}}]{brown2004}
{Brown}, M.~E., {Trujillo}, C., \& {Rabinowitz}, D. 2004, \apj, 617, 645

\bibitem[{{Burchell} {et~al.}(2005){Burchell}, {Leliwa-Kopysty{\'n}ski}, \&
  {Arakawa}}]{burchell2005}
{Burchell}, M.~J., {Leliwa-Kopysty{\'n}ski}, J., \& {Arakawa}, M. 2005,
  \icarus, 179, 274

\bibitem[{{Calvet} {et~al.}(2005){Calvet}, {D'Alessio}, {Watson},
  {Franco-Hern{\'a}ndez}, {Furlan}, {Green}, {Sutter}, {Forrest}, {Hartmann},
  {Uchida}, {Keller}, {Sargent}, {Najita}, {Herter}, {Barry}, \&
  {Hall}}]{calvet2005}
{Calvet}, N., {D'Alessio}, P., {Watson}, D.~M., {et~al.} 2005, \apjl, 630, L185

\bibitem[{{Canovas} {et~al.}(2016){Canovas}, {Caceres}, {Schreiber}, {Hardy},
  {Cieza}, {M{\'e}nard}, \& {Hales}}]{canovas2016}
{Canovas}, H., {Caceres}, C., {Schreiber}, M.~R., {et~al.} 2016, \mnras, 458,
  L29

\bibitem[{{Chatterjee} {et~al.}(2008){Chatterjee}, {Ford}, {Matsumura}, \&
  {Rasio}}]{chatterjee2008}
{Chatterjee}, S., {Ford}, E.~B., {Matsumura}, S., \& {Rasio}, F.~A. 2008, \apj,
  686, 580

\bibitem[{{Chavanis}(2000)}]{chavanis2000}
{Chavanis}, P.~H. 2000, \aap, 356, 1089

\bibitem[{{Chen} {et~al.}(2013){Chen}, {Kavelaars}, {Gwyn}, {Ferrarese},
  {C{\^o}t{\'e}}, {Jord{\'a}n}, {Suc}, {Cuillandre}, \& {Ip}}]{chen2013}
{Chen}, Y.-T., {Kavelaars}, J.~J., {Gwyn}, S., {et~al.} 2013, \apjl, 775, L8

\bibitem[{{Chiang} \& {Youdin}(2010)}]{chiang2010}
{Chiang}, E., \& {Youdin}, A.~N. 2010, Annual Review of Earth and Planetary
  Sciences, 38, 493

\bibitem[{{Cieza} {et~al.}(2010){Cieza}, {Schreiber}, {Romero}, {Mora},
  {Merin}, {Swift}, {Orellana}, {Williams}, {Harvey}, \& {Evans}}]{cieza2010}
{Cieza}, L.~A., {Schreiber}, M.~R., {Romero}, G.~A., {et~al.} 2010, \apj, 712,
  925

\bibitem[{{Clarke} {et~al.}(2001){Clarke}, {Gendrin}, \&
  {Sotomayor}}]{clarke2001}
{Clarke}, C.~J., {Gendrin}, A., \& {Sotomayor}, M. 2001, \mnras, 328, 485

\bibitem[{{Cloutier} {et~al.}(2014){Cloutier}, {Currie}, {Rieke}, {Kenyon},
  {Balog}, \& {Jayawardhana}}]{clout2014}
{Cloutier}, R., {Currie}, T., {Rieke}, G.~H., {et~al.} 2014, \apj, 796, 127

\bibitem[{{Corder} {et~al.}(2009){Corder}, {Carpenter}, {Sargent}, {Zauderer},
  {Wright}, {White}, {Woody}, {Teuben}, {Scott}, {Pound}, {Plambeck}, {Lamb},
  {Koda}, {Hodges}, {Hawkins}, \& {Bock}}]{corder2009}
{Corder}, S., {Carpenter}, J.~M., {Sargent}, A.~I., {et~al.} 2009, \apjl, 690,
  L65

\bibitem[{{Currie} {et~al.}(2014{\natexlab{a}}){Currie}, {Daemgen}, {Debes},
  {Lafreniere}, {Itoh}, {Jayawardhana}, {Ratzka}, \& {Correia}}]{currie2014a}
{Currie}, T., {Daemgen}, S., {Debes}, J., {et~al.} 2014{\natexlab{a}}, \apjl,
  780, L30

\bibitem[{{Currie} {et~al.}(2009){Currie}, {Lada}, {Plavchan}, {Robitaille},
  {Irwin}, \& {Kenyon}}]{currie2009}
{Currie}, T., {Lada}, C.~J., {Plavchan}, P., {et~al.} 2009, \apj, 698, 1

\bibitem[{{Currie} {et~al.}(2014{\natexlab{b}}){Currie}, {Muto}, {Kudo},
  {Honda}, {Brandt}, {Grady}, {Fukagawa}, {Burrows}, {Janson}, {Kuzuhara},
  {McElwain}, {Follette}, {Hashimoto}, {Henning}, {Kandori}, {Kusakabe},
  {Kwon}, {Mede}, {Morino}, {Nishikawa}, {Pyo}, {Serabyn}, {Suenaga},
  {Takahashi}, {Wisniewski}, \& {Tamura}}]{currie2014b}
{Currie}, T., {Muto}, T., {Kudo}, T., {et~al.} 2014{\natexlab{b}}, \apjl, 796,
  L30

\bibitem[{{Davis} {et~al.}(1985){Davis}, {Chapman}, {Weidenschilling}, \&
  {Greenberg}}]{davis1985}
{Davis}, D.~R., {Chapman}, C.~R., {Weidenschilling}, S.~J., \& {Greenberg}, R.
  1985, Icarus, 63, 30

\bibitem[{{de la Fuente Marcos} \& {de la Fuente Marcos}(2014)}]{delafm2014}
{de la Fuente Marcos}, C., \& {de la Fuente Marcos}, R. 2014, \mnras, 443, L59

\bibitem[{{Espaillat} {et~al.}(2010){Espaillat}, {D'Alessio}, {Hern{\'a}ndez},
  {Nagel}, {Luhman}, {Watson}, {Calvet}, {Muzerolle}, \&
  {McClure}}]{espaillat2010}
{Espaillat}, C., {D'Alessio}, P., {Hern{\'a}ndez}, J., {et~al.} 2010, \apj,
  717, 441

\bibitem[{{Fienga} {et~al.}(2016){Fienga}, {Laskar}, {Manche}, \&
  {Gastineau}}]{fienga2016}
{Fienga}, A., {Laskar}, J., {Manche}, H., \& {Gastineau}, M. 2016, ArXiv
  e-prints, arXiv:1602.06116

\bibitem[{{Ford} \& {Rasio}(2008)}]{ford2008}
{Ford}, E.~B., \& {Rasio}, F.~A. 2008, \apj, 686, 621

\bibitem[{{Forgan} \& {Rice}(2013)}]{forgan2013}
{Forgan}, D., \& {Rice}, K. 2013, \mnras, 432, 3168

\bibitem[{{Garaud} {et~al.}(2013){Garaud}, {Meru}, {Galvagni}, \&
  {Olczak}}]{garaud2013}
{Garaud}, P., {Meru}, F., {Galvagni}, M., \& {Olczak}, C. 2013, \apj, 764, 146

\bibitem[{{Gibbons} {et~al.}(2015){Gibbons}, {Mamatsashvili}, \&
  {Rice}}]{gibbons2015}
{Gibbons}, P.~G., {Mamatsashvili}, G.~R., \& {Rice}, W.~K.~M. 2015, \mnras,
  453, 4232

\bibitem[{{Giblin} {et~al.}(2004){Giblin}, {Davis}, \& {Ryan}}]{giblin2004}
{Giblin}, I., {Davis}, D.~R., \& {Ryan}, E.~V. 2004, Icarus, 171, 487

\bibitem[{{Gladman} {et~al.}(2009){Gladman}, {Kavelaars}, {Petit}, {Ashby},
  {Parker}, {Coffey}, {Jones}, {Rousselot}, \& {Mousis}}]{gladman2009}
{Gladman}, B., {Kavelaars}, J., {Petit}, J.-M., {et~al.} 2009, \apjl, 697, L91

\bibitem[{{Goldreich} {et~al.}(2004){Goldreich}, {Lithwick}, \&
  {Sari}}]{gold2004}
{Goldreich}, P., {Lithwick}, Y., \& {Sari}, R. 2004, \araa, 42, 549

\bibitem[{{Gomes} {et~al.}(2006){Gomes}, {Matese}, \& {Lissauer}}]{gomes2006}
{Gomes}, R.~S., {Matese}, J.~J., \& {Lissauer}, J.~J. 2006, \icarus, 184, 589

\bibitem[{{Gonzalez} {et~al.}(2015){Gonzalez}, {Laibe}, {Maddison}, {Pinte}, \&
  {M{\'e}nard}}]{gonzalez2015}
{Gonzalez}, J.-F., {Laibe}, G., {Maddison}, S.~T., {Pinte}, C., \&
  {M{\'e}nard}, F. 2015, \planss, 116, 48

\bibitem[{{Gorti} \& {Hollenbach}(2009)}]{gorti2009}
{Gorti}, U., \& {Hollenbach}, D. 2009, \apj, 690, 1539

\bibitem[{{Gorti} {et~al.}(2015){Gorti}, {Hollenbach}, \&
  {Dullemond}}]{gorti2015}
{Gorti}, U., {Hollenbach}, D., \& {Dullemond}, C.~P. 2015, \apj, 804, 29

\bibitem[{{Haghighipour} \& {Boss}(2003)}]{haghi2003}
{Haghighipour}, N., \& {Boss}, A.~P. 2003, \apj, 583, 996

\bibitem[{{Haisch} {et~al.}(2001){Haisch}, {Lada}, \& {Lada}}]{haisch2001}
{Haisch}, Jr., K.~E., {Lada}, E.~A., \& {Lada}, C.~J. 2001, \apjl, 553, L153

\bibitem[{{Hartmann} {et~al.}(1998){Hartmann}, {Calvet}, {Gullbring}, \&
  {D'Alessio}}]{hart1998}
{Hartmann}, L., {Calvet}, N., {Gullbring}, E., \& {D'Alessio}, P. 1998, \apj,
  495, 385

\bibitem[{{Haworth} {et~al.}(2016){Haworth}, {Clarke}, \& {Owen}}]{haworth2016}
{Haworth}, T.~J., {Clarke}, C.~J., \& {Owen}, J.~E. 2016, \mnras, 457, 1905

\bibitem[{{Helled} {et~al.}(2014){Helled}, {Bodenheimer}, {Podolak}, {Boley},
  {Meru}, {Nayakshin}, {Fortney}, {Mayer}, {Alibert}, \& {Boss}}]{helled2014}
{Helled}, R., {Bodenheimer}, P., {Podolak}, M., {et~al.} 2014, in Protostars
  and Planets VI, ed. {H.~Beuther, R.~S.~Klessen, C.~P.~Dullemond, \&
  T.~Henning} (University of Arizona Press, Tucson, AZ), 643--665

\bibitem[{{Heng} \& {Kenyon}(2010)}]{heng2010b}
{Heng}, K., \& {Kenyon}, S.~J. 2010, \mnras, 408, 1476

\bibitem[{{Holsapple}(1994)}]{hols1994}
{Holsapple}, K.~A. 1994, \planss, 42, 1067

\bibitem[{{Housen} \& {Holsapple}(1999)}]{housen1999}
{Housen}, K.~R., \& {Holsapple}, K.~A. 1999, Icarus, 142, 21

\bibitem[{{Ida} {et~al.}(2000){Ida}, {Larwood}, \& {Burkert}}]{ida2000a}
{Ida}, S., {Larwood}, J., \& {Burkert}, A. 2000, \apj, 528, 351

\bibitem[{{Iorio}(2012)}]{iorio2012}
{Iorio}, L. 2012, Celestial Mechanics and Dynamical Astronomy, 112, 117

\bibitem[{{Iorio}(2014)}]{iorio2014}
---. 2014, \mnras, 444, L78

\bibitem[{{J{\'{\i}}lkov{\'a}} {et~al.}(2015){J{\'{\i}}lkov{\'a}}, {Portegies
  Zwart}, {Pijloo}, \& {Hammer}}]{jilkova2015}
{J{\'{\i}}lkov{\'a}}, L., {Portegies Zwart}, S., {Pijloo}, T., \& {Hammer}, M.
  2015, \mnras, 453, 3157

\bibitem[{{Johansen} {et~al.}(2015){Johansen}, {Mac Low}, {Lacerda}, \&
  {Bizzarro}}]{johan2015}
{Johansen}, A., {Mac Low}, M.-M., {Lacerda}, P., \& {Bizzarro}, M. 2015,
  Science Advances, 1, 15109

\bibitem[{{Kenyon} \& {Bromley}(2001)}]{kb2001}
{Kenyon}, S.~J., \& {Bromley}, B.~C. 2001, \aj, 121, 538

\bibitem[{{Kenyon} \& {Bromley}(2002)}]{kb2002b}
---. 2002, \apjl, 577, L35

\bibitem[{{Kenyon} \& {Bromley}(2004{\natexlab{a}})}]{kb2004a}
---. 2004{\natexlab{a}}, \aj, 127, 513

\bibitem[{{Kenyon} \& {Bromley}(2004{\natexlab{b}})}]{kb2004b}
---. 2004{\natexlab{b}}, \apjl, 602, L133

\bibitem[{{Kenyon} \& {Bromley}(2004{\natexlab{c}})}]{kb2004d}
---. 2004{\natexlab{c}}, \nat, 432, 598

\bibitem[{{Kenyon} \& {Bromley}(2004{\natexlab{d}})}]{kb2004c}
---. 2004{\natexlab{d}}, \aj, 128, 1916

\bibitem[{{Kenyon} \& {Bromley}(2006)}]{kb2006}
---. 2006, \aj, 131, 1837

\bibitem[{{Kenyon} \& {Bromley}(2008)}]{kb2008}
---. 2008, \apjs, 179, 451

\bibitem[{{Kenyon} \& {Bromley}(2010)}]{kb2010}
---. 2010, \apjs, 188, 242

\bibitem[{{Kenyon} \& {Bromley}(2012)}]{kb2012}
---. 2012, \aj, 143, 63

\bibitem[{{Kenyon} \& {Bromley}(2015)}]{kb2015a}
---. 2015, \apj, 806, 42

\bibitem[{{Kenyon} \& {Bromley}(2016)}]{kb2016}
---. 2016, \apj, 817, 51

\bibitem[{{Kenyon} {et~al.}(2008){Kenyon}, {G{\'o}mez}, \& {Whitney}}]{kgw2008}
{Kenyon}, S.~J., {G{\'o}mez}, M., \& {Whitney}, B.~A. 2008, in Handbook of Star
  Forming Regions, Volume I, ed. {Reipurth, B.}, 405--458

\bibitem[{{Kenyon} \& {Hartmann}(1995)}]{kh1995}
{Kenyon}, S.~J., \& {Hartmann}, L. 1995, \apjs, 101, 117

\bibitem[{{Kenyon} \& {Luu}(1998)}]{kl1998}
{Kenyon}, S.~J., \& {Luu}, J.~X. 1998, \aj, 115, 2136

\bibitem[{{Kenyon} \& {Luu}(1999)}]{kl1999a}
---. 1999, \aj, 118, 1101

\bibitem[{{Kim} {et~al.}(2013){Kim}, {Watson}, {Manoj}, {Forrest}, {Najita},
  {Furlan}, {Sargent}, {Espaillat}, {Muzerolle}, {Megeath}, {Calvet}, {Green},
  \& {Arnold}}]{kim2013}
{Kim}, K.~H., {Watson}, D.~M., {Manoj}, P., {et~al.} 2013, \apj, 769, 149

\bibitem[{{Klahr} \& {Henning}(1997)}]{klahr1997}
{Klahr}, H.~H., \& {Henning}, T. 1997, \icarus, 128, 213

\bibitem[{{Kobayashi} \& {Tanaka}(2010)}]{koba2010a}
{Kobayashi}, H., \& {Tanaka}, H. 2010, \icarus, 206, 735

\bibitem[{{Kobayashi} {et~al.}(2010){Kobayashi}, {Tanaka}, {Krivov}, \&
  {Inaba}}]{koba2010b}
{Kobayashi}, H., {Tanaka}, H., {Krivov}, A.~V., \& {Inaba}, S. 2010, \icarus,
  209, 836

\bibitem[{{Krist} {et~al.}(2012){Krist}, {Stapelfeldt}, {Bryden}, \&
  {Plavchan}}]{krist2012}
{Krist}, J.~E., {Stapelfeldt}, K.~R., {Bryden}, G., \& {Plavchan}, P. 2012,
  \aj, 144, 45

\bibitem[{{Krist} {et~al.}(2010){Krist}, {Stapelfeldt}, {Bryden}, {Rieke},
  {Su}, {Chen}, {Beichman}, {Hines}, {Rebull}, {Tanner}, {Trilling}, {Clampin},
  \& {G{\'a}sp{\'a}r}}]{krist2010}
{Krist}, J.~E., {Stapelfeldt}, K.~R., {Bryden}, G., {et~al.} 2010, \aj, 140,
  1051

\bibitem[{{Lafreni{\`e}re} {et~al.}(2010){Lafreni{\`e}re}, {Jayawardhana}, \&
  {van Kerkwijk}}]{lafren2010}
{Lafreni{\`e}re}, D., {Jayawardhana}, R., \& {van Kerkwijk}, M.~H. 2010, \apj,
  719, 497

\bibitem[{{Lagrange} {et~al.}(2010){Lagrange}, {Bonnefoy}, {Chauvin}, {Apai},
  {Ehrenreich}, {Boccaletti}, {Gratadour}, {Rouan}, {Mouillet}, {Lacour}, \&
  {Kasper}}]{lagrange2010}
{Lagrange}, A.-M., {Bonnefoy}, M., {Chauvin}, G., {et~al.} 2010, Science, 329,
  57

\bibitem[{{Leinhardt} \& {Stewart}(2009)}]{lein2009}
{Leinhardt}, Z.~M., \& {Stewart}, S.~T. 2009, \icarus, 199, 542

\bibitem[{{Leinhardt} \& {Stewart}(2012)}]{lein2012}
---. 2012, \apj, 745, 79

\bibitem[{{Leinhardt} {et~al.}(2008){Leinhardt}, {Stewart}, \&
  {Schultz}}]{lein2008}
{Leinhardt}, Z.~M., {Stewart}, S.~T., \& {Schultz}, P.~H. 2008, in The Solar
  System Beyond Neptune, ed. {Barucci, M.~A., Boehnhardt, H., Cruikshank,
  D.~P., \& Morbidelli, A.} (University of Arizona Press, Tucson, AZ), 195--211

\bibitem[{{Levison} {et~al.}(2010){Levison}, {Duncan}, {Brasser}, \&
  {Kaufmann}}]{levison2010b}
{Levison}, H.~F., {Duncan}, M.~J., {Brasser}, R., \& {Kaufmann}, D.~E. 2010,
  Science, 329, 187

\bibitem[{{Li} \& {Adams}(2016)}]{li2016}
{Li}, G., \& {Adams}, F.~C. 2016, ArXiv e-prints, arXiv:1602.08496

\bibitem[{{Lissauer}(1987)}]{liss1987}
{Lissauer}, J.~J. 1987, Icarus, 69, 249

\bibitem[{{Love} \& {Ahrens}(1996)}]{love1996}
{Love}, S.~G., \& {Ahrens}, T.~J. 1996, Icarus, 124, 141

\bibitem[{{Lykawka} \& {Mukai}(2008)}]{lykaw2008}
{Lykawka}, P.~S., \& {Mukai}, T. 2008, \aj, 135, 1161

\bibitem[{{Madigan} \& {McCourt}(2016)}]{madigan2016}
{Madigan}, A.-M., \& {McCourt}, M. 2016, \mnras, 457, L89

\bibitem[{{Mamajek}(2009)}]{mama2009}
{Mamajek}, E.~E. 2009, in American Institute of Physics Conference Series, Vol.
  1158, American Institute of Physics Conference Series, ed. {T.~Usuda,
  M.~Tamura, \& M.~Ishii}, 3--10

\bibitem[{{Marois} {et~al.}(2008){Marois}, {Macintosh}, {Barman}, {Zuckerman},
  {Song}, {Patience}, {Lafreni{\`e}re}, \& {Doyon}}]{marois2008}
{Marois}, C., {Macintosh}, B., {Barman}, T., {et~al.} 2008, Science, 322, 1348

\bibitem[{{Marshall} {et~al.}(2011){Marshall}, {L{\"o}hne}, {Montesinos},
  {Krivov}, {Eiroa}, {Absil}, {Bryden}, {Maldonado}, {Mora}, {Sanz-Forcada},
  {Ardila}, {Augereau}, {Bayo}, {Del Burgo}, {Danchi}, {Ertel}, {Fedele},
  {Fridlund}, {Lebreton}, {Gonz{\'a}lez-Garc{\'{\i}}a}, {Liseau}, {Meeus},
  {M{\"u}ller}, {Pilbratt}, {Roberge}, {Stapelfeldt}, {Th{\'e}bault}, {White},
  \& {Wolf}}]{marshall2011}
{Marshall}, J.~P., {L{\"o}hne}, T., {Montesinos}, B., {et~al.} 2011, \aap, 529,
  A117

\bibitem[{{Marzari} {et~al.}(2010){Marzari}, {Baruteau}, \&
  {Scholl}}]{marzari2010}
{Marzari}, F., {Baruteau}, C., \& {Scholl}, H. 2010, \aap, 514, L4

\bibitem[{{Matese} {et~al.}(1999){Matese}, {Whitman}, \&
  {Whitmire}}]{matese1999}
{Matese}, J.~J., {Whitman}, P.~G., \& {Whitmire}, D.~P. 1999, \icarus, 141, 354

\bibitem[{{Michikoshi} {et~al.}(2007){Michikoshi}, {Inutsuka}, {Kokubo}, \&
  {Furuya}}]{michi2007}
{Michikoshi}, S., {Inutsuka}, S.-i., {Kokubo}, E., \& {Furuya}, I. 2007, \apj,
  657, 521

\bibitem[{{Michikoshi} {et~al.}(2009){Michikoshi}, {Kokubo}, \&
  {Inutsuka}}]{michi2009}
{Michikoshi}, S., {Kokubo}, E., \& {Inutsuka}, S.-i. 2009, \apj, 703, 1363

\bibitem[{{Michikoshi} {et~al.}(2010){Michikoshi}, {Kokubo}, \&
  {Inutsuka}}]{michi2010}
---. 2010, \apj, 719, 1021

\bibitem[{{Moeckel} \& {Armitage}(2012)}]{moeckel2012}
{Moeckel}, N., \& {Armitage}, P.~J. 2012, \mnras, 419, 366

\bibitem[{{Moeckel} {et~al.}(2008){Moeckel}, {Raymond}, \&
  {Armitage}}]{moeckel2008}
{Moeckel}, N., {Raymond}, S.~N., \& {Armitage}, P.~J. 2008, \apj, 688, 1361

\bibitem[{{Morbidelli} \& {Levison}(2004{\natexlab{a}})}]{morby2004a}
{Morbidelli}, A., \& {Levison}, H.~F. 2004{\natexlab{a}}, \aj, 128, 2564

\bibitem[{{Morbidelli} \& {Levison}(2004{\natexlab{b}})}]{morbi2004}
---. 2004{\natexlab{b}}, \aj, 128, 2564

\bibitem[{{Nagasawa} \& {Ida}(2011)}]{naga2011}
{Nagasawa}, M., \& {Ida}, S. 2011, \apj, 742, 72

\bibitem[{{Najita} {et~al.}(2015){Najita}, {Andrews}, \&
  {Muzerolle}}]{najita2015}
{Najita}, J.~R., {Andrews}, S.~M., \& {Muzerolle}, J. 2015, \mnras, 450, 3559

\bibitem[{{Najita} \& {Kenyon}(2014)}]{najita2014}
{Najita}, J.~R., \& {Kenyon}, S.~J. 2014, \mnras, 445, 3315

\bibitem[{{Najita} {et~al.}(2007){Najita}, {Strom}, \&
  {Muzerolle}}]{najita2007b}
{Najita}, J.~R., {Strom}, S.~E., \& {Muzerolle}, J. 2007, \mnras, 378, 369

\bibitem[{{Nayakshin}(2010)}]{nayak2010}
{Nayakshin}, S. 2010, \mnras, 408, L36

\bibitem[{{Nayakshin}(2015)}]{nayak2015}
---. 2015, \mnras, 454, 64

\bibitem[{{O'Brien} \& {Greenberg}(2003)}]{obrien2003}
{O'Brien}, D.~P., \& {Greenberg}, R. 2003, Icarus, 164, 334

\bibitem[{{Ohtsuki}(1999)}]{oht1999}
{Ohtsuki}, K. 1999, \icarus, 137, 152

\bibitem[{{Ohtsuki} {et~al.}(2002){Ohtsuki}, {Stewart}, \& {Ida}}]{oht2002}
{Ohtsuki}, K., {Stewart}, G.~R., \& {Ida}, S. 2002, Icarus, 155, 436

\bibitem[{{Owen} {et~al.}(2012){Owen}, {Clarke}, \& {Ercolano}}]{owen2012}
{Owen}, J.~E., {Clarke}, C.~J., \& {Ercolano}, B. 2012, \mnras, 422, 1880

\bibitem[{{Owen} {et~al.}(2013){Owen}, {Hudoba de Badyn}, {Clarke}, \&
  {Robins}}]{owen2013}
{Owen}, J.~E., {Hudoba de Badyn}, M., {Clarke}, C.~J., \& {Robins}, L. 2013,
  \mnras, 436, 1430

\bibitem[{{Rabinowitz} {et~al.}(2013){Rabinowitz}, {Schwamb}, {Hadjiyska},
  {Tourtellotte}, \& {Rojo}}]{rabin2013}
{Rabinowitz}, D., {Schwamb}, M.~E., {Hadjiyska}, E., {Tourtellotte}, S., \&
  {Rojo}, P. 2013, \aj, 146, 17

\bibitem[{{Raettig} {et~al.}(2015){Raettig}, {Klahr}, \& {Lyra}}]{raettig2015}
{Raettig}, N., {Klahr}, H., \& {Lyra}, W. 2015, \apj, 804, 35

\bibitem[{{Rafikov}(2004)}]{raf2004}
{Rafikov}, R.~R. 2004, \aj, 128, 1348

\bibitem[{{Rasio} \& {Ford}(1996)}]{rasio1996}
{Rasio}, F.~A., \& {Ford}, E.~B. 1996, Science, 274, 954

\bibitem[{{Ricci} {et~al.}(2015){Ricci}, {Carpenter}, {Fu}, {Hughes}, {Corder},
  \& {Isella}}]{ricci2015}
{Ricci}, L., {Carpenter}, J.~M., {Fu}, B., {et~al.} 2015, \apj, 798, 124

\bibitem[{{Rice}(2016)}]{rice2016}
{Rice}, K. 2016, ArXiv e-prints, arXiv:1602.08390

\bibitem[{{Ryan} {et~al.}(1999){Ryan}, {Davis}, \& {Giblin}}]{ryan1999}
{Ryan}, E.~V., {Davis}, D.~R., \& {Giblin}, I. 1999, Icarus, 142, 56

\bibitem[{{Sheppard}(2010)}]{sheppard2010a}
{Sheppard}, S.~S. 2010, \aj, 139, 1394

\bibitem[{{Simon} {et~al.}(2015){Simon}, {Armitage}, {Li}, \&
  {Youdin}}]{simon2015}
{Simon}, J.~B., {Armitage}, P.~J., {Li}, R., \& {Youdin}, A.~N. 2015, ArXiv
  e-prints, arXiv:1512.00009

\bibitem[{{Stern} \& {Colwell}(1997)}]{stcol1997a}
{Stern}, S.~A., \& {Colwell}, J.~E. 1997, \aj, 114, 841

\bibitem[{{Tanaka} {et~al.}(1996){Tanaka}, {Inaba}, \&
  {Nakazawa}}]{tanaka1996b}
{Tanaka}, H., {Inaba}, S., \& {Nakazawa}, K. 1996, Icarus, 123, 450

\bibitem[{{Tobin} {et~al.}(2015){Tobin}, {Looney}, {Wilner}, {Kwon},
  {Chandler}, {Bourke}, {Loinard}, {Chiang}, {Schnee}, \& {Chen}}]{tobin2015}
{Tobin}, J.~J., {Looney}, L.~W., {Wilner}, D.~J., {et~al.} 2015, \apj, 805, 125

\bibitem[{{Trujillo} \& {Sheppard}(2014)}]{trujillo2014}
{Trujillo}, C.~A., \& {Sheppard}, S.~S. 2014, \nat, 507, 471

\bibitem[{{Weidenschilling}(1977)}]{weiden1977a}
{Weidenschilling}, S.~J. 1977, \mnras, 180, 57

\bibitem[{{Weidenschilling}(1989)}]{weiden1989}
---. 1989, Icarus, 80, 179

\bibitem[{{Weidenschilling}(2010)}]{weid2010}
---. 2010, \apj, 722, 1716

\bibitem[{{Weidenschilling} \& {Marzari}(1996)}]{weiden1996}
{Weidenschilling}, S.~J., \& {Marzari}, F. 1996, \nat, 384, 619

\bibitem[{{Wetherill} \& {Stewart}(1993)}]{weth1993}
{Wetherill}, G.~W., \& {Stewart}, G.~R. 1993, Icarus, 106, 190

\bibitem[{{Williams} \& {Wetherill}(1994)}]{will1994}
{Williams}, D.~R., \& {Wetherill}, G.~W. 1994, Icarus, 107, 117

\bibitem[{{Williams} \& {Cieza}(2011)}]{will2011}
{Williams}, J.~P., \& {Cieza}, L.~A. 2011, \araa, 49, 67

\bibitem[{{Williams} {et~al.}(2004){Williams}, {Najita}, {Liu}, {Bottinelli},
  {Carpenter}, {Hillenbrand}, {Meyer}, \& {Soderblom}}]{will2004}
{Williams}, J.~P., {Najita}, J., {Liu}, M.~C., {et~al.} 2004, \apj, 604, 414

\bibitem[{{Windmark} {et~al.}(2012){Windmark}, {Birnstiel}, {Ormel}, \&
  {Dullemond}}]{windmark2012}
{Windmark}, F., {Birnstiel}, T., {Ormel}, C.~W., \& {Dullemond}, C.~P. 2012,
  \aap, 544, L16

\bibitem[{{Youdin} \& {Chiang}(2004)}]{youdin2004a}
{Youdin}, A.~N., \& {Chiang}, E.~I. 2004, \apj, 601, 1109

\bibitem[{{Youdin} \& {Kenyon}(2013)}]{youdin2013}
{Youdin}, A.~N., \& {Kenyon}, S.~J. 2013, {From Disks to Planets}, ed. T.~D.
  {Oswalt}, L.~M. {French}, \& P.~{Kalas} (Dordrecht: Springer Science \&
  Business Media), 1

\bibitem[{{Youdin} \& {Shu}(2002)}]{youdin2002}
{Youdin}, A.~N., \& {Shu}, F.~H. 2002, \apj, 580, 494

\bibitem[{{Zhu} {et~al.}(2014){Zhu}, {Stone}, {Rafikov}, \& {Bai}}]{zhu2014}
{Zhu}, Z., {Stone}, J.~M., {Rafikov}, R.~R., \& {Bai}, X.-N. 2014, \apj, 785,
  122

\end{thebibliography}

\clearpage
\begin{deluxetable}{ll}
\tablecolumns{7}
\tablewidth{0pc}
\tabletypesize{\small}
\tablenum{1}
\tablecaption{List of Variables }
\tablehead{
\colhead{Variable} &
\colhead{Definition}
}
\startdata
$a$ & semimajor axis, radial coordinate \\
$a_{in}$ & semimajor axis of the inner edge of the disk \\
$a_{max}$ & semimajor axis of the maximum pressure in the disk \\
$a_{out}$ & semimajor axis of the outer edge of the disk \\
$\ad$ & cross-sectional area of particles \\
$\bd$ & exponent in relation for debris from collisions \\
$\bl$ & exponent in relation for mass of largest particle in debris \\
$e$ & eccentricity \\
$e_{rel}$ & eccentricity relative to the Hill radius of the largest object \\
$f_g$ & gravitational focusing factor \\
$f_0$ & fraction of initial mass in oligarchs \\
$h$ & horizontal velocity \\
$i$ & inclination \\
$L_d$ & reprocessed stellar luminosity of solid particles \\
\lstar\ & stellar luminosity \\
$m$, $m_k$ & particle mass \\
$\bar{n}$ & average mass of particle in a mass bin \\
$\mmaxd$ & mass of largest particle in debris \\
$\ml0$ & coefficient in relation for mass of largest particle in debris \\
$m_{esc}$ & mass of debris ejected in a collision \\
$m_{max}$ & mass of largest particle in the grid\\
$m_{min}$ & mass of smallest particle in the grid\\
$M_0$ & total initial mass in particles \\
\mstar\ & stellar mass \\
$N$, $N_k$ & particle number \\
$N_{max}$ & number of largest particles \\
$P$ & gas pressure in the disk \\
$P_{max}$ & maximum gas pressure in the disk \\
$q_p$ & perihelion distance \\
$Q_b$, $Q_g$ & coefficients in \qdstar\ relation \\
$Q_c$ & center of mass collision energy \\
$\qdstar$ & collision energy required to eject 50\% of the mass \\
$r$, $r_k$ & particle radius \\
$\bar{r}$ & average radius of particle in a mass bin \\
$r_{max}$ & radius of largest particle \\
$r_{min}$ & radius of smallest particle \\
$\rstar $ & radius of central star \\
$t$ & time \\
$T$ & temperature \\
$v$ & vertical velocity \\
$v_c$ & relative collision velocity \\
$v_K$ & orbital velocity \\
$V$ & volume \\
$\beta_b$, $\beta_g$ & exponents in \qdstar\ relation \\
$\delta$ & mass spacing factor \\
$\delta a$ & width of annulus \\
$\rho_p$ & particle mass density \\
$\sigma$ & geometric cross section \\
$\Sigma$ & surface density \\
$\omega$ & argument of perihelion \\
$\Omega$ & angular velocity \\
$\varpi$ & longitude of perihelion \\
\enddata
\tablecomments{Variables with a subscript `0' refer to initial conditions;
e.g., $e_0$ is the initial eccentricity}
\label{tab: pars}
\end{deluxetable}
\clearpage

%
\clearpage

\clearpage

\clearpage

\clearpage

\clearpage

\clearpage

\clearpage

\clearpage

\end{document}